\documentclass[useAMS,usenatbib]{mn2e}
\usepackage{graphicx}
\usepackage{latexsym}
\usepackage{psfig}
\usepackage{amssymb,amsmath}
\usepackage{subfig}
\usepackage{caption}

\title{Early evolution of embedded clusters}

\author[J. E. Dale, B. Ercolano, I.A. Bonnell]{J. E. Dale$^{1,2}$\thanks{E-mail: dale@usm.lmu.de (JED)}, B. Ercolano$^{1,2}$, I. A. Bonnell$^{3}$\\
$^{1}$Excellence Cluster `Universe', Boltzmannstr. 2, 85748 Garching, Germany.\\
$^{2}$Universit\"{a}ts--Sternwarte M\"{u}nchen, Scheinerstr. 1, 81679 M\"{u}nchen, Germany.\\
$^{3}$Department of Physics and Astronomy, University of St Andrews, North Haugh, St Andrews, Fife KY16 9SS}

\begin{document}

\pagerange{\pageref{firstpage}--\pageref{lastpage}} \pubyear{2006}

\maketitle


\def\mnras{MNRAS}
\def\apj{ApJ}
\def\aj{AJ}
\def\aap{A\&A}
\def\apjl{ApJL}
\def\apjs{ApJS}
\def\araa{ARA\&A}
\def\pasp{PASP}
 
\begin{abstract}
We examine the combined effects of winds and photoionizing radiation from O--type stars on embedded stellar clusters formed in model turbulent molecular clouds covering a range of masses and radii. We find that feedback is able to increase the quantities of dense gas present, but decreases the rate and efficiency of the conversion of gas to stars relative to control simulations in which feedback is absent. Star formation in these calculations often proceeds at a rate substantially slower than the freefall rate in the dense gas. This decoupling is due to the weakening of, and expulsion of gas from, the deepest parts of the clouds' potential wells where most of the star formation occurs in the control simulations. This results in large fractions of the stellar populations in the feedback simulation becoming dissociated from dense gas. However, where star formation \emph{does} occur in both control and feedback simulations, it does so in dense gas, so the correlation between star formation activity and dense gas is preserved.\\
\indent The overall dynamical effects of feedback on the \emph{clusters} are minimal, with only small fraction of stars becoming unbound, despite large quantities of gas being expelled from some clouds. This owes to the settling of the stars into virialised and stellar--dominated configurations before the onset of feedback. By contrast, the effects of feedback on the observable properties of the clusters -- their U--, B-- and V--band magnitudes -- are strong and sudden. The timescales on which the clusters become visible and unobscured are short compared with the timescales which the clouds are actually destroyed.\\
\end{abstract}

\begin{keywords}
stars: formation
\end{keywords}
\section{Introduction}
\indent Star formation occurs inside giant molecular clouds (GMCs) in an hierarchical fashion. At the largest scales, clouds convert a small fraction, typically a few percent, of their gas reservoir to stars before dispersing on timescales of $\sim10$Myr. Most stellar groupings dissolve into the field on a similar timescale, but several classes of identifiable objects, such as open clusters, OB associations, scaled OB associations and globular clusters, survive for much longer times. What it is that governs which of these paths a given young stellar population takes is not clear.\\
\indent There are many processes acting on different timescales which can disperse a group of stars into the field, recently reviewed and discussed by \cite{2012MNRAS.426.3008K}. The mechanism most intimately connected with the star formation process itself, and acting on the shortest timescales, is referred to as `infant mortality' or `infant weight--loss' \citep[][]{2003ARA&A..41...57L,2006MNRAS.369L...9B}. These authors gathered together observations of embedded clusters and showed that the rate at which embedded clusters are formed is 10--20 times higher than what would be expected from the rate at which gas--free open clusters are formed, if all embedded clusters were to evolve into open clusters. They therefore inferred that $\sim$90$\%$ of embedded clusters do not evolve to become open clusters, but are instead dispersed on the same timescale (5--10Myr) as that on which they evolve from the embedded to the exposed or open phase. This implies that the expulsion of gas that makes the clusters visible is also what destroys them.\\
\indent \cite{2005ApJ...631L.133F} and \cite{2007AJ....133.1067W} use the large number of clusters visible in the Antennae galaxies to plot mass--age diagrams in which they also deduced that $\sim90\%$ of clusters are lost in each logarithmic age bin, but that the cluster mass function was unaffected. The dispersal mechanisms acting on each timescale seem therefore to be mass--independent.\\
\indent This issue was revisited by \cite{2008A&A...482..165G} who instead examined the most massive clusters in logarithmic age bins in the Antennae, M51,  the SMC and LMC, M33 and M83. Their results for the Antennae were in agreement with those of \cite{2005ApJ...631L.133F} and \cite{2007AJ....133.1067W}, and they also found evidence of mass--independent cluster disruption in M51, but not in the other systems studied. \cite{2009ApJ...701..607B} corrected for the non--constancy of star formation in the Antennae merger system and found evidence for cluster disruption taking place on timescales $<10$Myr, but not on longer timescales. The observational picture of clusters at very young ages is thus somewhat murky.\\
\indent From the theoretical point of view, there are two potential ways of ensuring that a recently--formed cluster disperses on a similar timescale to that on which star formation goes to completion and the cluster becomes exposed. The first is gas expulsion and this has been extensively investigated. Most such studies have relied on analytic models or N--body simulations where the gas is represented, in one way or another, as an external potential which is removed over some timescale \citep[e.g][]{1978A&A....70...57T,1980ApJ...235..986H,1984ApJ...285..141L,2003MNRAS.338..665B,2003MNRAS.338..673B,2006MNRAS.373..752G,2007MNRAS.380.1589B,2013A&A...555A.135P}. These simulations have generally found that gas removal while the potential is still gas--dominated unbinds large fractions of the stars. However, hydrodynamic simulations call into question whether the potential in which the stars are situated is ever gas--dominated \citep[e.g.][]{2009ApJ...704L.124O,2012MNRAS.419..841K,2012MNRAS.420..613G}.\\
\indent With improvements in algorithms and increases in computing power, it has become possible to model both star formation and stellar feedback self--consistently in hydrodynamic simulations. Some authors have concentrated on the effects of thermal accretion feedback \citep[e.g][]{2009MNRAS.392.1363B,2010ApJ...710.1343U} or jets/outflows from low--mass stars \citep[e.g.][]{2006ApJ...640L.187L,2011ApJ...740..107C,2012ApJ...754...71K,2014ApJ...790..128F,2014MNRAS.439.3420M}. These works have generally concentrated on the effects of feedback on setting the masses of individual stars, the determination of the stellar initial mass function, and the setting of the overall star formation efficiency. These feedback processes are expected to have limited influence the large--scale dynamics of the clouds or the clusters, except insofar as they reduce the rate at which gas is converted to stars.\\
{\indent At larger scales, modelling of the destructive effects of HII regions on whole GMCs has been popular, since an HII region can in principle expand and clear out a large fraction of the volume of a molecular cloud on a relative short timescale. These simulations have produced somewhat mixed results, depending on the clouds under investigation. \cite{2012MNRAS.427..625W} modelled the effect of ionising radiation from central O--stars on clouds with various fractal dimensions and find it to be highly destructive, dispersing their 6.4pc radius, 10$^{4}$M$_{\odot}$ clouds in a few Myr. Similar disruption timescales in clouds formed by colliding flows were reported by \cite{2013MNRAS.435.1701C}.} However, \cite{2005MNRAS.358..291D}, who modelled ionising radiation emanating from an O--star at the hub of a network of filamentary accretion flows, \cite{2010ApJ...711.1017P} (modelling irradiation from O--stars forming near the centre of a massive disk) and \cite{2011MNRAS.414..321D} (simulating the influence of massive stars on a 10$^{6}$M$_{\odot}$ cloud with a high escape velocity) all observed their HII regions to flicker as they were swamped by neutral gas delivered by accretion flows. In these calculations, the dynamical influence of photoionisation on the scales of whole clouds was much more modest. \cite{2012MNRAS.424..377D} and \cite{2013MNRAS.430..234D} examined this issue in a suite of simulations spanning a GMC mass--radius parameter space ranging from 10$^{4}$--10$^{6}$M$_{\odot}$ in mass and 5--180 pc in radius. They confirmed that accretion flows restrict the ability of HII regions to disrupt clouds. However, they also showed that the clouds' escape velocities were crucially important (as predicted by \cite{2002ApJ...566..302M} for example), since HII regions cannot expand at speeds much exceeding the (roughly constant) sound speed in gas photoinised by O--stars of $\approx10$km s$^{-1}$.\\
\indent \cite{2014arXiv1410.0011W} investigated the combined influence of photoionisation and supernova explosions, building on the work of \cite{2012MNRAS.427..625W}. They found that the effect of the photoionisation phase before the supernova explosion was to delay somewhat the transition of the supernova remnant from the Sedov--Taylor phase to the radiative phase, by reducing the density of the gas encountered by the remnant. This allows the supernova to deposit $\approx50\%$ more momentum into the cold gas.\\
\indent Instead of photoionisation, \cite{2012MNRAS.420.1503P} model the effects of winds and supernova explosions from massive stars on an embedded cluster. They do not model star formation, but they do investigate the influence of feedback on an admixture of stars and gas. They find that the influence of feedback depends very strongly on the efficiency with which the clouds retain the injected energy (which they choose to parameterise), but that star formation efficiencies as low as 5$\%$ can result in bound systems surviving gas expulsion.\\
\indent The second possible explanation for the apparent poor survival chances of embedded clusters, investigated by \cite{2005MNRAS.359..809C}, is that the clusters are never bound in the first place. This seems plausible on the face of it, since GMCs exist with a variety of virial ratios \citep[e.g][]{2011MNRAS.413.2935D}. However, two problems exist. Firstly, star formation in unbound clouds tends to result in flat stellar mass functions \citep{2008MNRAS.386....3C}. Secondly, since the stars are able to decouple from the gas dynamics at early ages, the cloud being unbound does not guarantee that the stars will be so.\\
\indent As a counterpoint to the foregoing discussion, there has been a resurgence of the idea that the space distribution of stars should more properly be thought of as \emph{hierachical}, rather than merely \emph{clustered} \citep[e.g][]{2001AJ....121.1507E, 2003MNRAS.343..413B,2007MNRAS.379.1302B}. This is partly driven by the difficulty of defining observationally what is and is not a cluster. \cite{2010MNRAS.409L..54B} highlighted this particularly strongly in their survey of the surface density of nearby YSO's in which they were unable to identify any distinctive scales which could be used for such a definition. While \cite{2012MNRAS.426L..11G} showed that this result does not necessarily imply that stars are not formed in bound clusters, it does show that observationally defining and identifying clusters is non--trivial. \cite{2007MNRAS.379.1302B} obtained similar results at much larger size--scales in their study of M33, in which they were unable to identify any preferred size scale for young stellar systems. We do not discuss our simulations in this context (Parker \& Dale, 2015, in preparation).\\
\indent This paper forms part of a series of studies of the effects of photoionisation and/or winds from O--type stars on a parameter space of model GMCs constructed to reflect the gross properties of the Milky Way clouds documented in \cite{2009ApJ...699.1092H}. In the first six papers, we performed controlled experiments involving ionisation \emph{or} winds with the object of disentangling their individual effects.\\
\indent In \cite{2014MNRAS.442..694D}, we combined the two forms of feedback and computed quantities global to our model clouds, such as the star formation efficiency, average star formation rate and unbound gas mass, to see how they varied across the parameter space. The results of the study can be briefly summarised as follows: The additional effect of winds on the dynamics of the cold gas was minimal, although the structure of the ionised gas was strongly altered in many of the simulations, being compressed by the winds into a thin shell lining the inner walls of the feedback--blown bubbles. The overall influence of feedback was strongly dependent on the clouds' escape velocities, with the lower--mass 10$^{4}$M$_{\odot}$ clouds having large fractions of their gas reserves unbound or expelled, whereas the 10$^{6}$M$_{\odot}$ clouds were much less severely damaged. Star formation rates and efficiencies followed a similar trend, being reduced by factors up to $\approx 2$ in the low--mass clouds, but scarcely changing in the larger objects. Feedback left the clouds substantially permeable to ionised gas, ionising photons and supernova debris.\\
\indent We previously focussed largely on the gas content of the clouds, mentioning the stars only as sources of feedback and sinks of gas. In this paper, we redress the balance by considering in more detail the properties of the stellar populations and clusters formed by the clouds. We examine the interplay of stars and gas in more detail and compare our results with more nuanced observational diagnostics than blunt instruments such as global star formation efficiencies.\\
\indent This paper proceeds as follows: Section 2 contains a brief recapitulation of our numerical methods. Section 3 describes the results derived from our simulations, and our discussion and conclusions follow in Sections 4 and 5.\\
\section{Numerical methods}
{We have investigated the formation of embedded clusters under the influence of stellar feedback using a set of Smoothed Particle Hydrodynamics (SPH) simulations of turbulent GMCs covering a mass--radius parameter space running from 10$^{4}$--10$^{6}$M$_{\odot}$ in mass and 2.5--180pc in radius. The clouds initially have shallow Gaussian density profiles, with the density contrast between the centres and edges of the clouds being approximately three. The clouds were given divergence--free turbulent velocity fields satisfying the scaling relation $P(k)\propto k^{-4}$, scaled to give them initial virial ratios of either 0.7 (which we refer to as `bound clouds') or 2.3 (which we refer to as `unbound clouds').}\\
\indent The thermal properties of the neutral gas are governed by a piecewise \cite{2005MNRAS.359..211L} equation of state defined by $P=k\rho^{\gamma}$ with a density--dependent adiabatic exponent given by
\begin{eqnarray}
\begin{array}{rlrl}
\gamma  &=  0.75  ; & \hfill &\rho \le \rho_1 \\
\gamma  &=  1.0  ; & \rho_1 \le & \rho  \le \rho_2 \\
\gamma  &=  1.4  ; & \hfill \rho_2 \le &\rho \le \rho_3 \\
\gamma  &=  1.0  ; & \hfill &\rho \ge \rho_3, \\
\end{array}
\label{eqn:eos}
\end{eqnarray}
and $\rho_1= 5.5 \times 10^{-19} {\rm g\ cm}^{-3} , \rho_2=5.5 \times10^{-15} {\rm g cm}^{-3} , \rho_3=2 \times 10^{-13} {\rm g\ cm}^{-3}$. The initial properties of the clouds are given in Table \ref{tab:sims}.\\
\begin{table*}
\begin{tabular}{|l|l|l|l|l|l|l|l|l|}
Run&Mass (M$_{\odot}$)&R$_{0}$(pc)&$\langle n(H_{2})\rangle$ (cm$^{-3}$) & v$_{\rm RMS,0}$(km s$^{-1}$)& $\langle T_{0}\rangle(K)$&$\langle {\mathcal M}_{0}\rangle$&t$_{\rm ff,0}$ (Myr)\\
\hline
A&$10^{6}$&180&2.9&5.0&143&6.5&19.6\\
\hline
B&$10^{6}$&95&16&6.9&92&11.2&7.50\\
\hline
X&$10^{6}$&45&149&9.6&52&20.7&2.44\\
\hline
D&$10^{5}$&45&15&3.0&92&4.9&7.70\\
\hline
E&$10^{5}$&21&147&4.6&52&4.9&2.46\\
\hline
F&$10^{5}$&10&1439&6.7&30&19.0&0.81\\
\hline
I&$10^{4}$&10&136&2.1&52&4.6&2.56\\
\hline
J&$10^{4}$&5&1135&3.0&30&8.5&0.90\\
\hline
\hline
UZ&$10^{6}$&45&149&18.2&52&39.3&2.9\\
\hline
UB&$3\times10^{5}$&45&45&10.0&68&18.9&6.0\\
\hline
UC&$3\times10^{5}$&21&443&14.6&36&37.9&1.9\\
\hline
UV&$10^{5}$&21&148&12.2&52&26.3&3.3\\
\hline
UU&$10^{5}$&10&1371&8.4&26&25.6&1.1\\
\hline
UF&$3\times10^{4}$&10&410&6.7&28&19.7&2.0\\
\hline
UP&$10^{4}$&2.5&9096&7.6&18&27.9&0.4\\
\hline
UQ&$10^{4}$&5.0&1137&5.4&30&14.9&1.2\\
\hline
\end{tabular}
\caption{Initial properties of clouds listed in descending order by mass. Columns are the run name, cloud mass, initial radius, initial RMS turbulent velocity, the initial mean gas temperature, the initial mean turbulent Mach number, and the initial cloud freefall time.}
\label{tab:sims}
\end{table*}
\indent The clouds are allowed to evolve, with the turbulence freely decaying, until they have formed a few massive stars or a few subclusters massive enough to host such stars. Stars and subclusters are represented by sink particles. The mass resolution of the $10^{4}$ M$_{\odot}$ and $3\times10^{4}$ M$_{\odot}$ clouds are 1 and 3 $M_{\odot}$ respectively, and sinks are taken to represent stars. Their accretion radii are $5\times10^{-3}$pc and their minimum creation densities are set to $7\times10^{7}$cm$^{-3}$. In the $10^{5}$ M$_{\odot}$, $3\times10^{5}$ M$_{\odot}$ and $10^{6}$ M$_{\odot}$ clouds, the mass resolutions are 10, 30, and 100$M_{\odot}$, and the sinks represent subclusters. Their accretion radii are set to either 0.1 or 0.25pc, whichever is less than 1 percent of the host cloud's initial radius. The minimum creation density of these objects is set to $4\times10^{5}$cm$^{-3}$ We then enable photoionisation and wind feedback using the algorithms presented in \cite{2007MNRAS.382.1759D,2012MNRAS.424..377D} and \cite{2008MNRAS.391....2D}.  In the $10^{4}$ M$_{\odot}$ and $3\times10^{4}$ M$_{\odot}$ clouds, sinks above $20$M$_{\odot}$ are assigned an ionising luminosity according to 
\begin{eqnarray}
{\rm log}(Q_{\rm H})=48.1+0.02(M_{*}-20M_{\odot}).
\end{eqnarray}
For the more massive clouds, sinks are treated as small clusters with Salpeter mass functions between 0.5 and 100M$_{\odot}$. The total mass contained in stars more massive than $30$M$_{\odot}$, M$_{30}$ is computed and the ionising luminosity of the sinks is set to (M$_{30}/30)\times Q_{30}$ s$^{-1}$ with $Q_{30}$ being the ionising luminosity of a 30M$_{\odot}$ star from the above equation.\\
\indent Wind momentum fluxes are set in a similar manner from mass loss rates given by
\begin{eqnarray}
\dot{M}(M_{*})=\left[0.3{\rm ~exp}\left(\frac{M_{*}}{28}\right)-0.3\right]\times10^{-6}{\rm M}_{\odot}{\rm yr}^{-1},
\label{eqn:mdot}
\end{eqnarray} 
and terminal velocities by
\begin{eqnarray}
v_{\infty}(M_{*})=\left[10^{3}(M_{*}-18)^{0.24}+600\right]{\rm km~s}^{-1}.
\label{eqn:vinf}
\end{eqnarray}
The clouds are then evolved for a further 3Myr (or as close to this as possible) to establish the combined effects of ionisation and winds in the interval between the formation of the first O--stars, and the first supernova explosions.\\
\indent The global effects of pre--supernova feedback were discussed in \cite{2014MNRAS.442..694D}. Here, we focus more on the detailed properties of the stellar clusters and the smaller--scale interplay between stars and gas with reference to recent observations.\\
\begin{figure*}
     \centering
   \subfloat[Run I]{\includegraphics[width=0.33\textwidth]{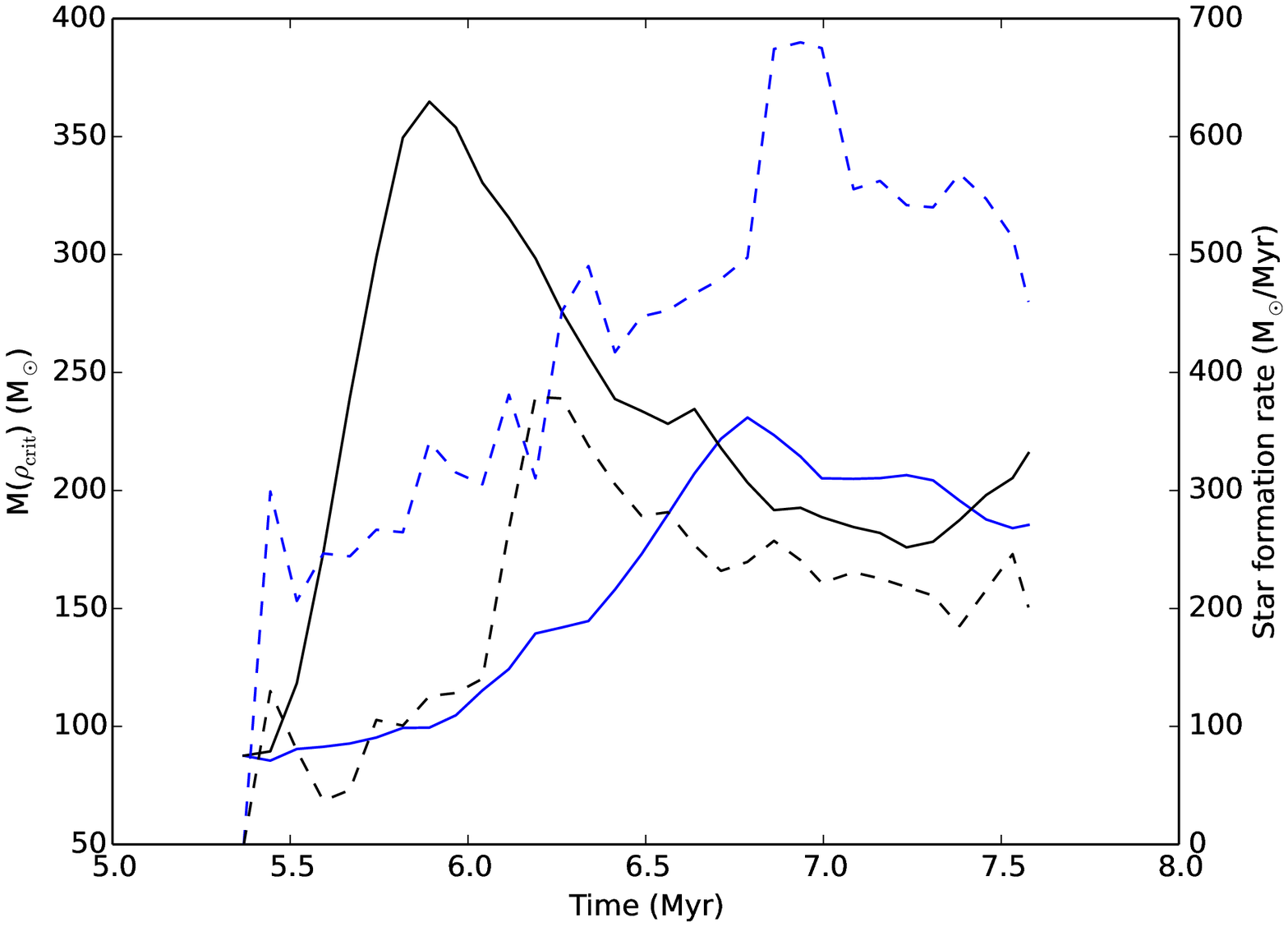}}
   \hspace{-0.05in}
      \subfloat[Run J]{\includegraphics[width=0.33\textwidth]{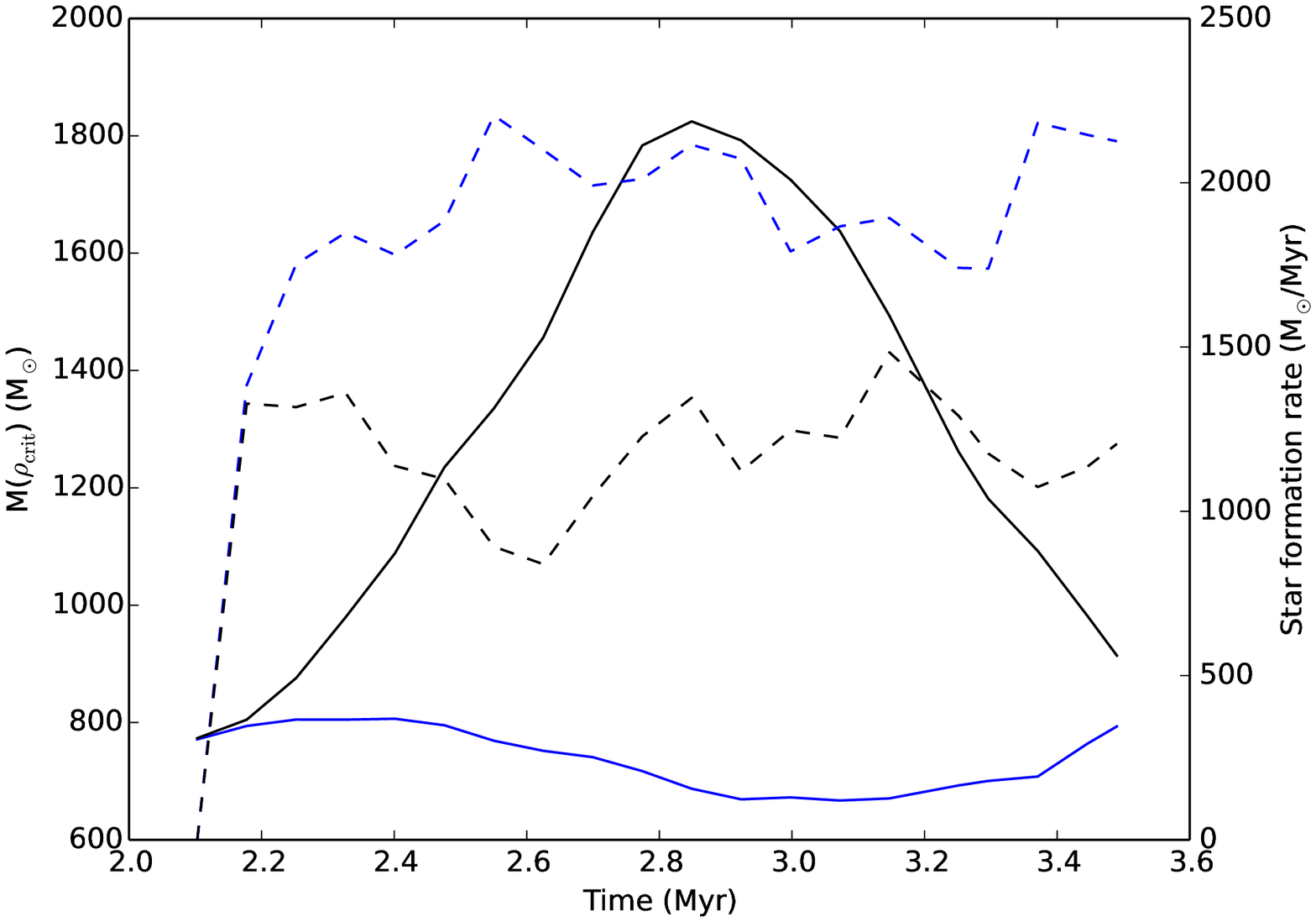}}
         \vspace{-0.03in}
      \subfloat[Run UF]{\includegraphics[width=0.33\textwidth]{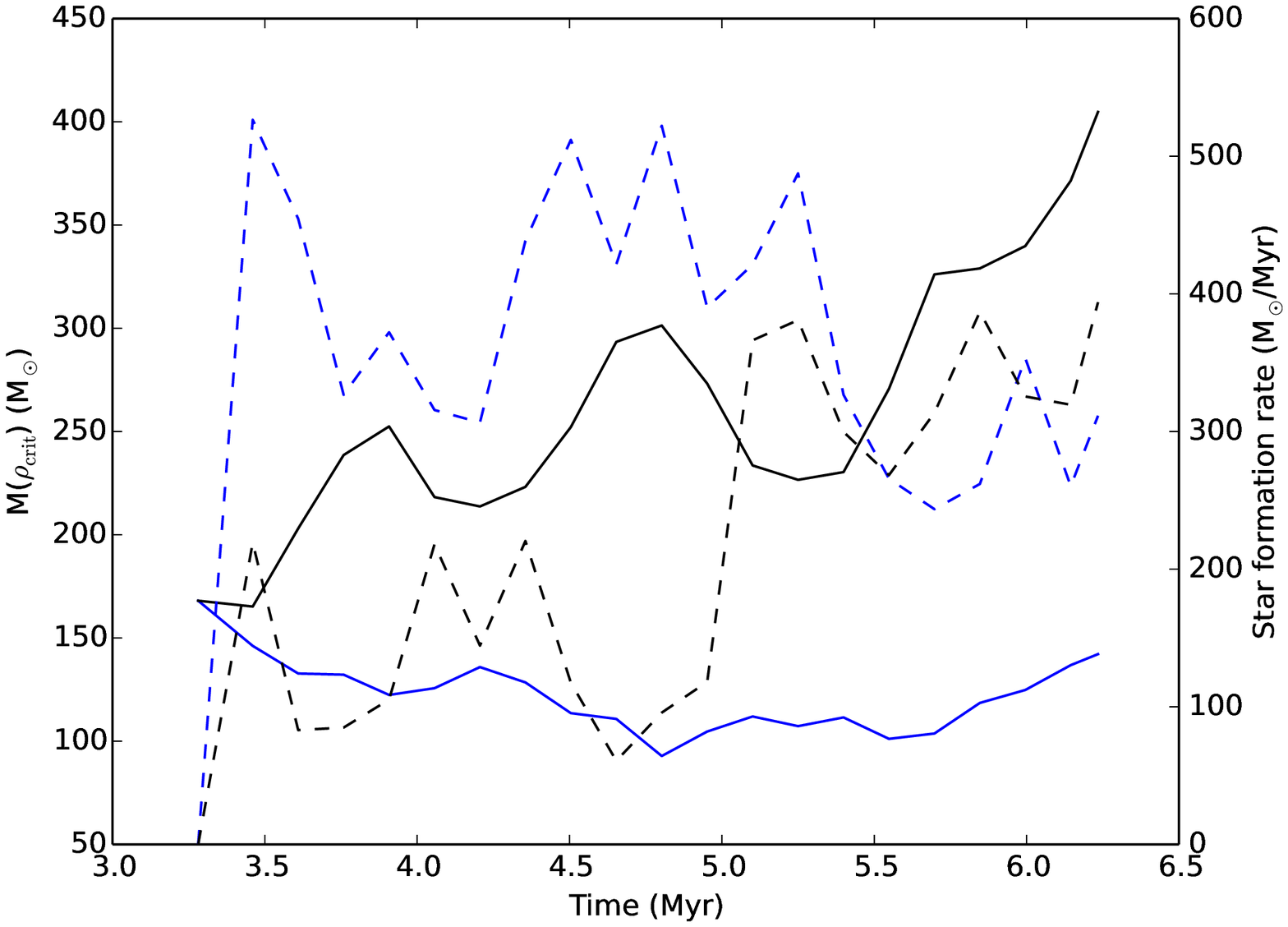}}
   \hspace{-0.05in}
      \subfloat[Run UP]{\includegraphics[width=0.33\textwidth]{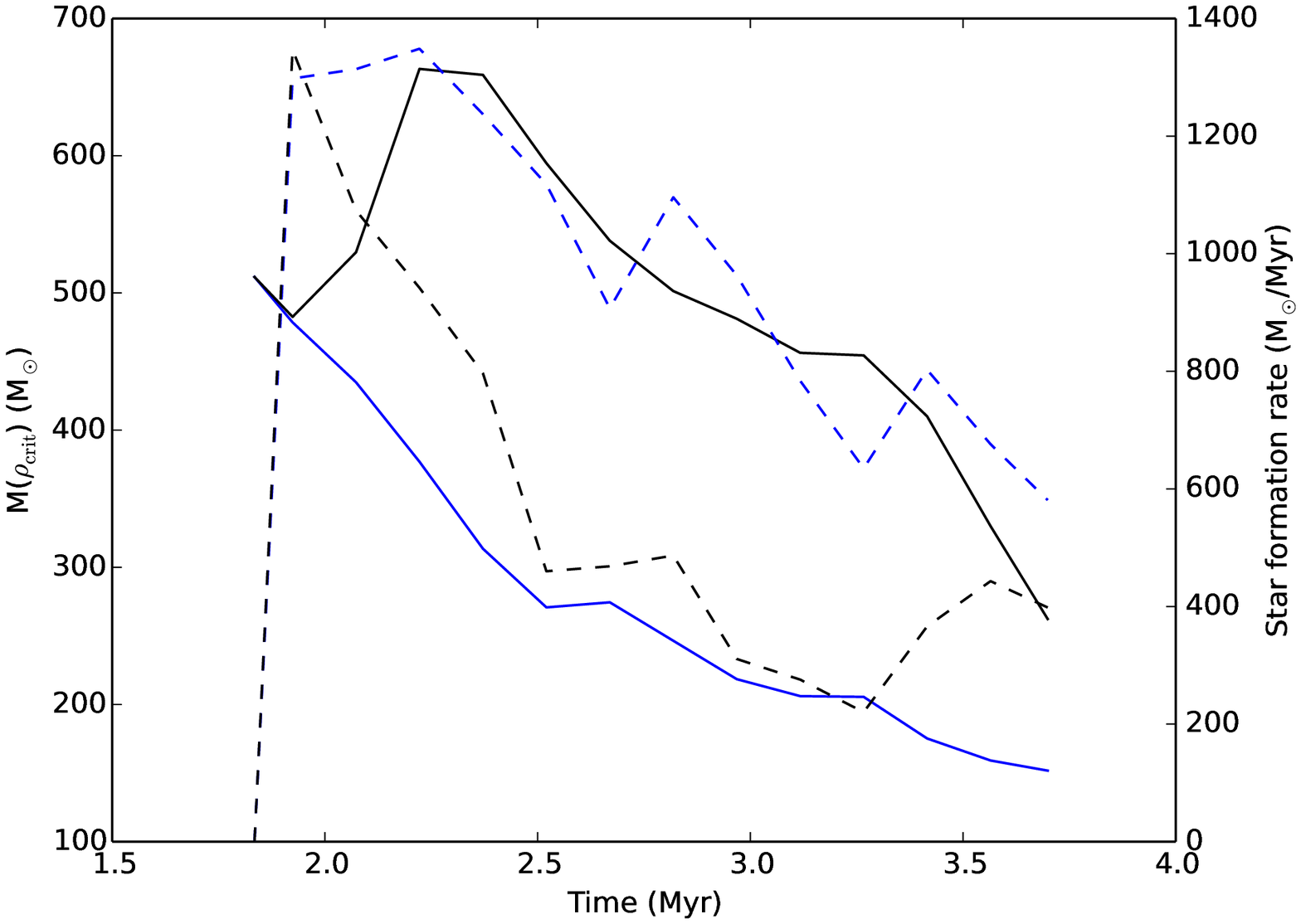}}
   \hspace{-0.05in}
      \subfloat[Run UQ]{\includegraphics[width=0.33\textwidth]{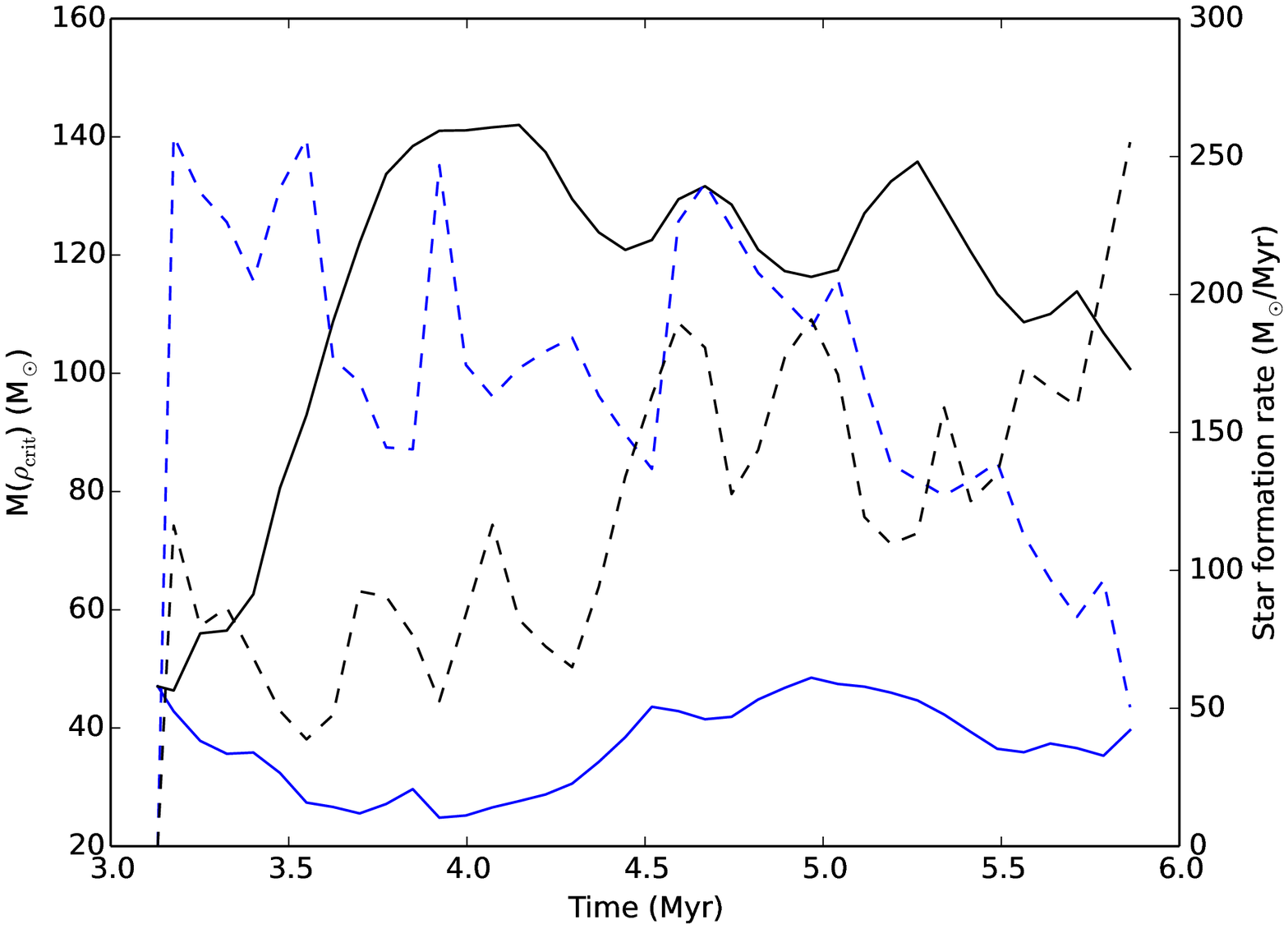}}
\caption{Comparison of the dense gas mass (left axis and solid lines) and star formation rates (right axis and dashed lines) in the control (blue) and dual--feedback (black) Runs I, J, UF, UP and UQ simulations.}
\label{fig:dens_sfr}
\end{figure*}
\begin{figure*}
     \centering
   \subfloat[Run I]{\includegraphics[width=0.33\textwidth]{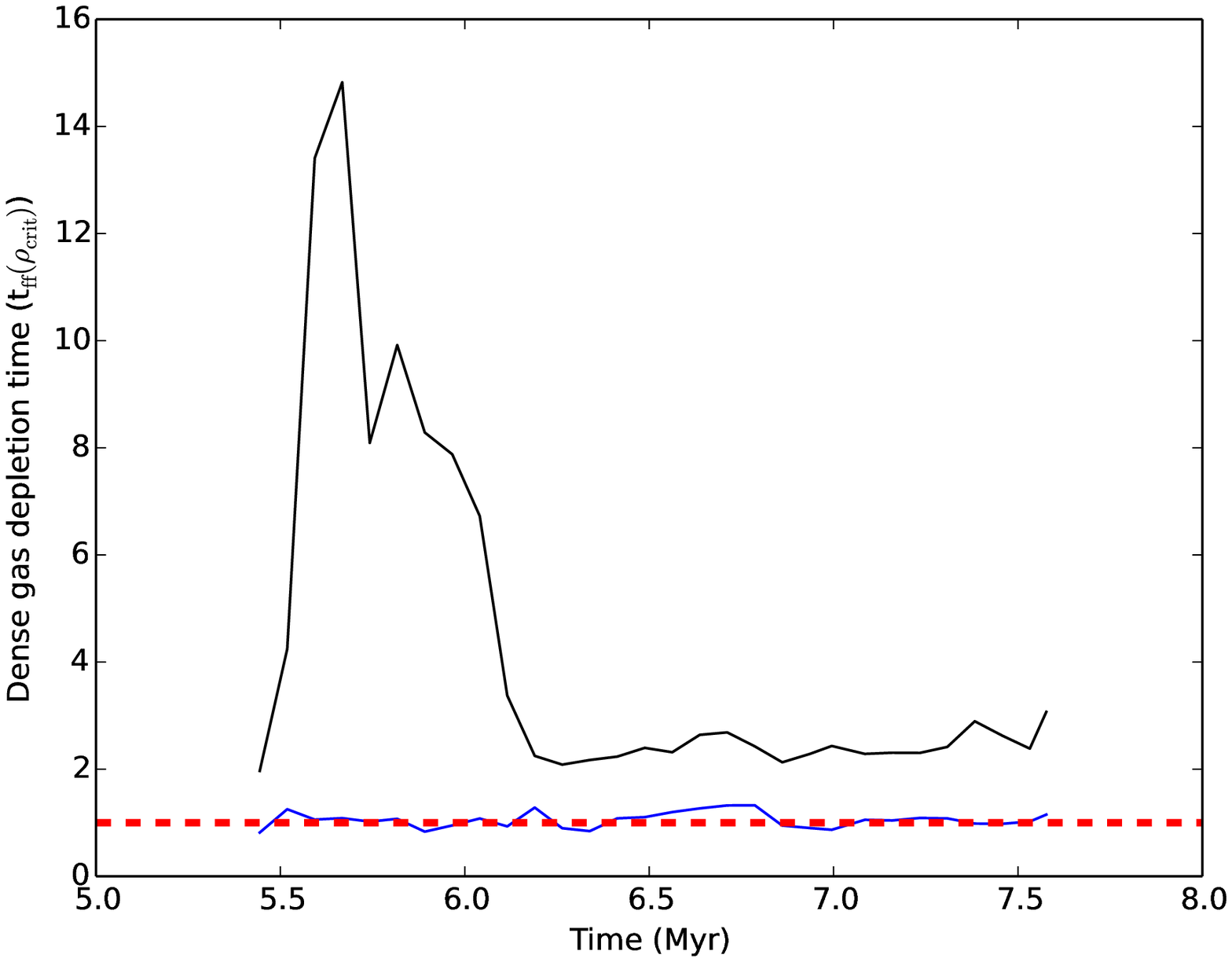}}
   \hspace{-0.05in}
      \subfloat[Run J]{\includegraphics[width=0.33\textwidth]{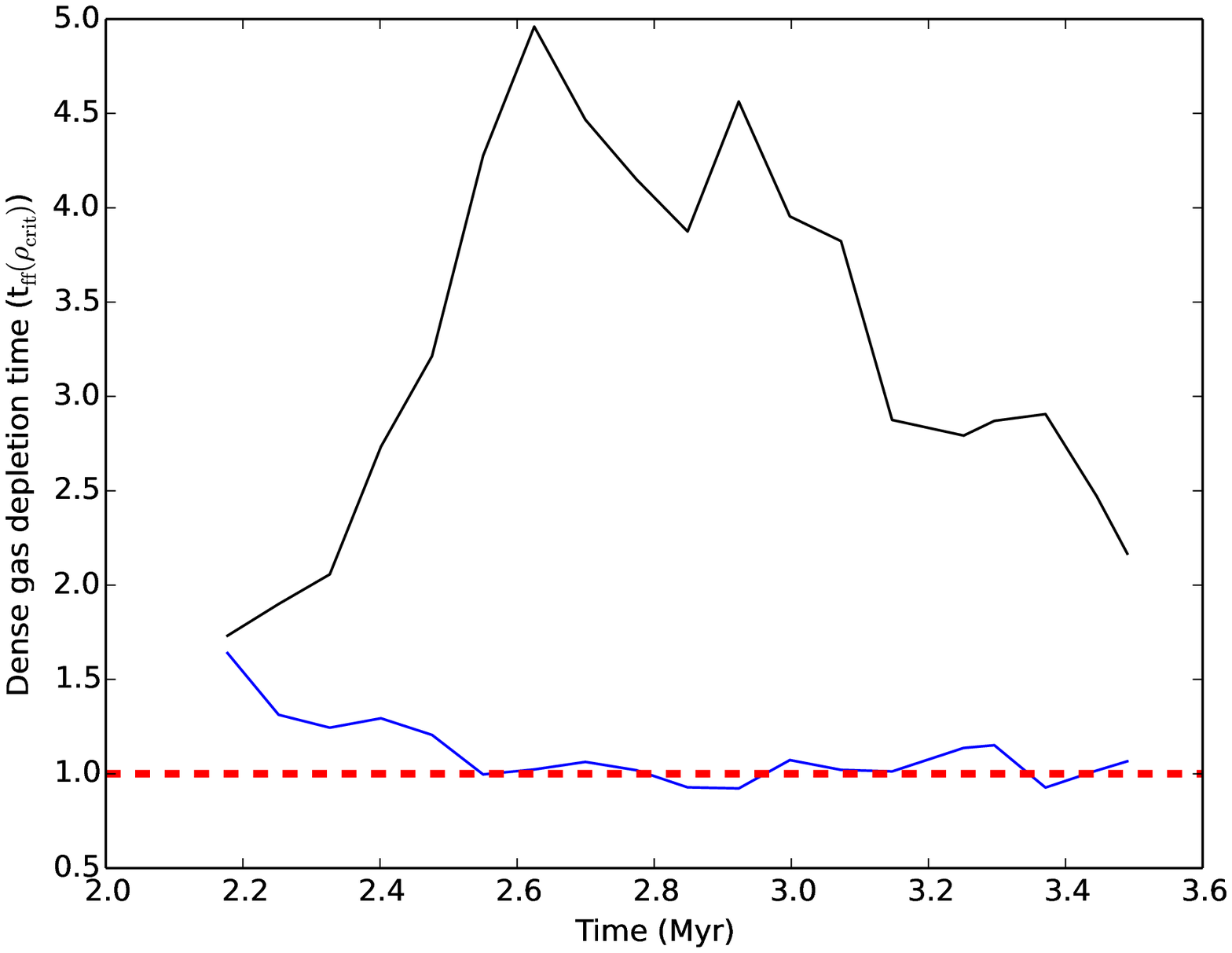}}
         \vspace{-0.03in}
      \subfloat[Run UF]{\includegraphics[width=0.33\textwidth]{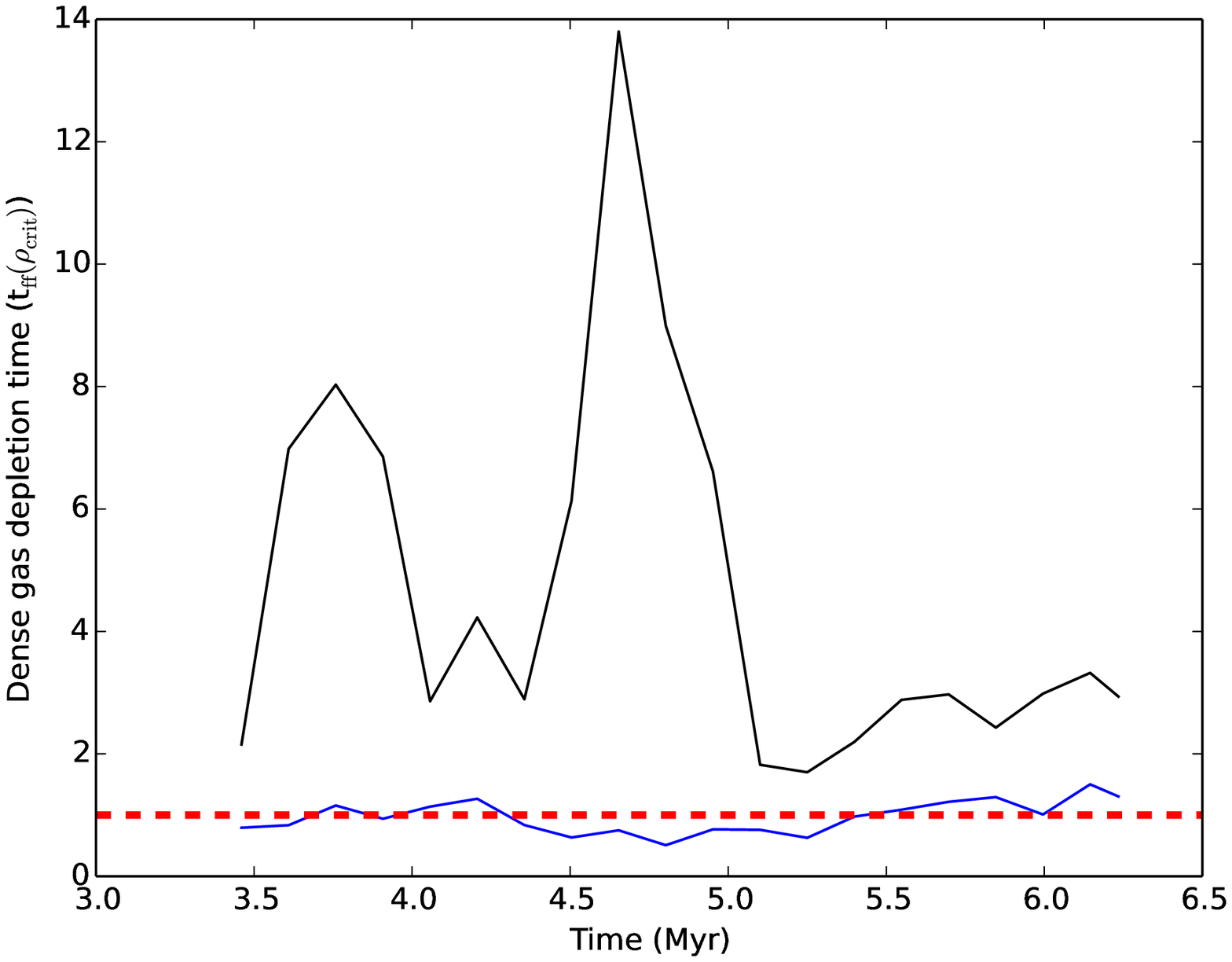}}
   \hspace{-0.05in}
      \subfloat[Run UP]{\includegraphics[width=0.33\textwidth]{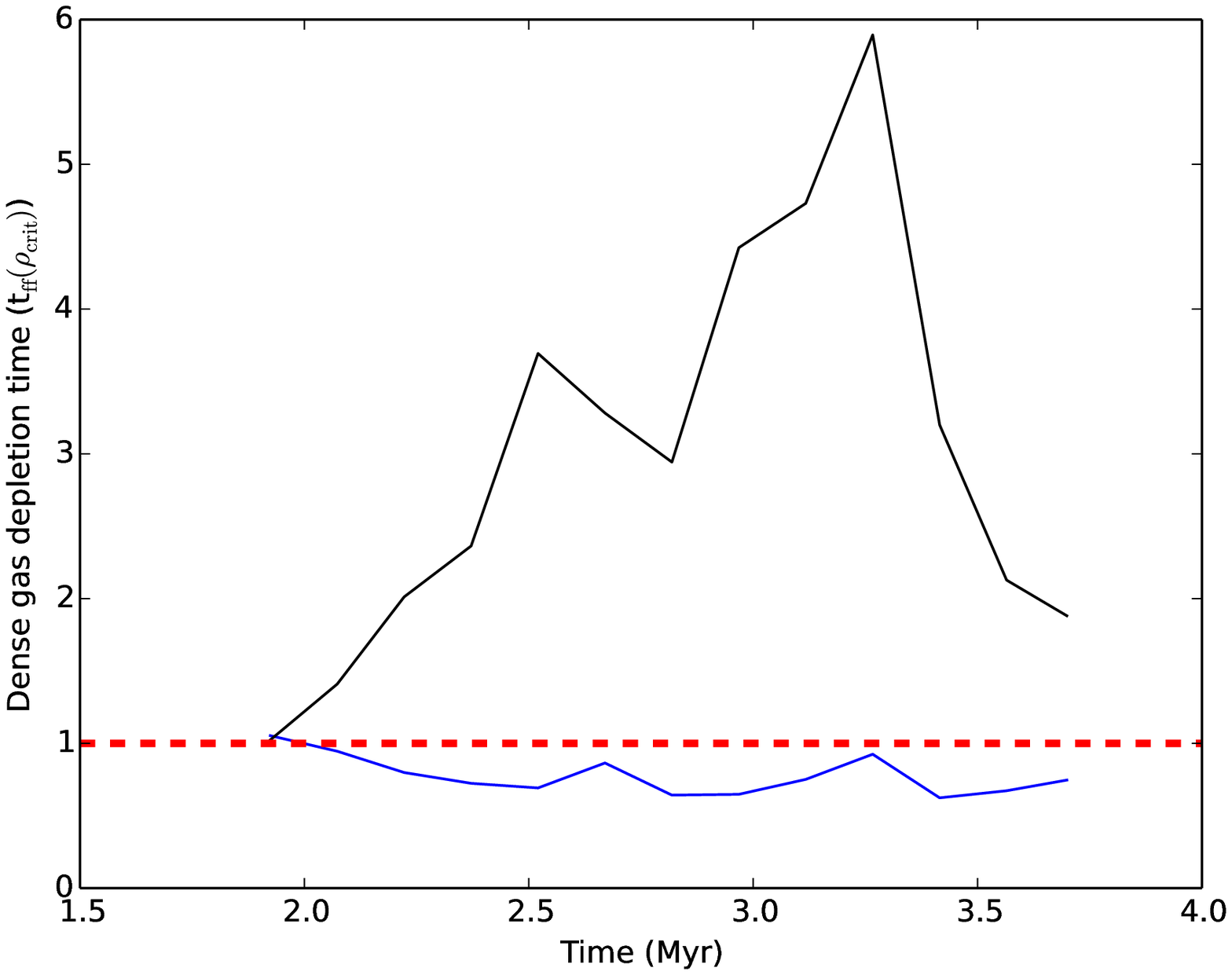}}
   \hspace{-0.05in}
      \subfloat[Run UQ]{\includegraphics[width=0.33\textwidth]{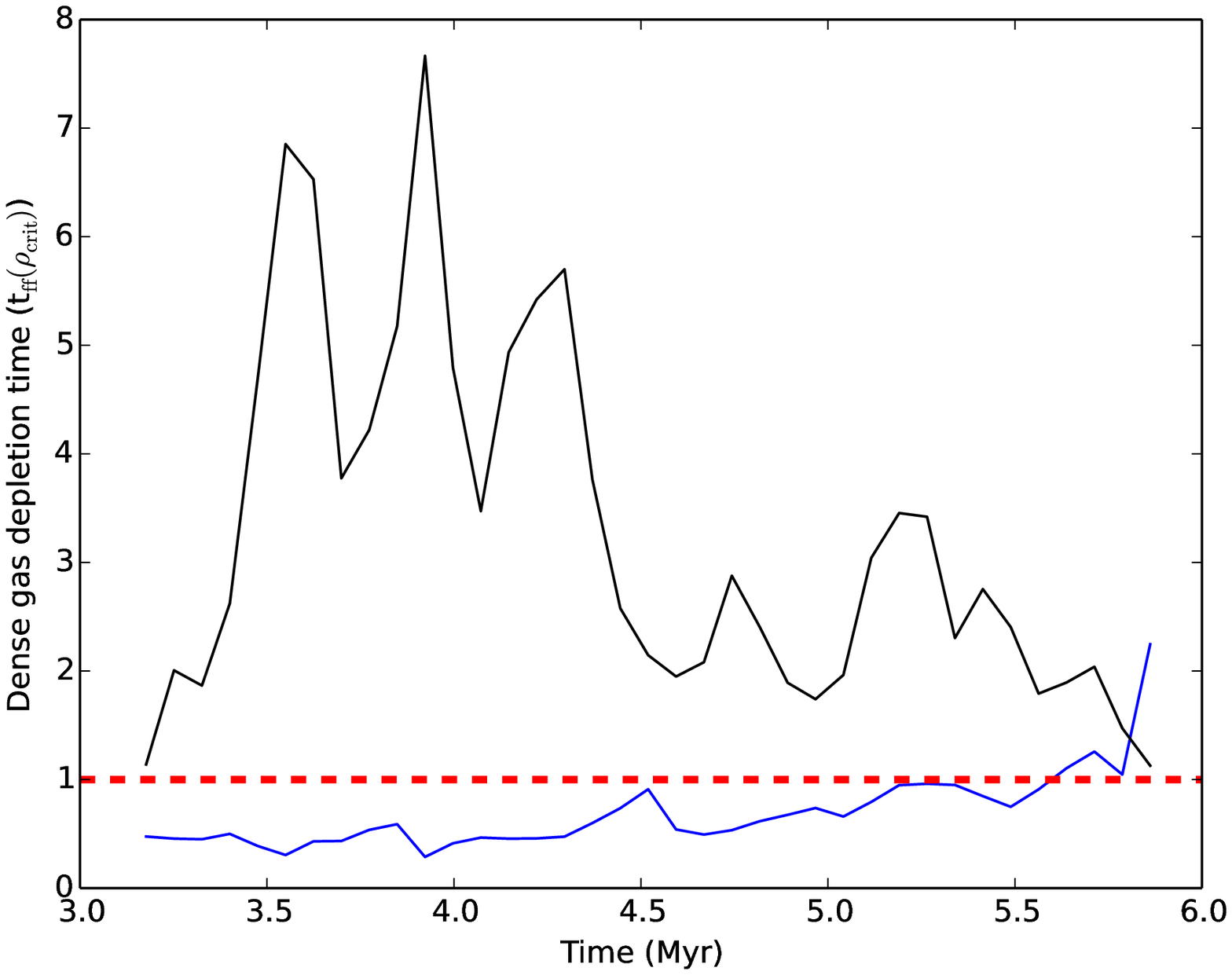}}
\caption{Dense gas mass depletion times in units of the freefall time at $\rho_{\rm crit}$ ($\approx 0.35$\,Myr) in the control (blue) and dual--feedback (black) Runs I, J, UF, UP and UQ simulations. The red dashed line corresponds to a depletion time of $t_{\rm ff}(\rho_{\rm crit}$), which would indicate star formation proceeding at the freefall rate in gas of this density.}
\label{fig:deptimes}
\end{figure*}
\section{Results}
\subsection{Dense gas depletion times}
\indent \cite{2010ApJ...724..687L} infer a linear relationship between the SFR and the mass of gas above a density threshold $\rho_{\rm crit}$ of 10$^{4}$ cm$^{-3}$. We examine this relationship in our simulations in Figure \ref{fig:dens_sfr} where we plot $M(\rho_{\rm crit})$ as a function of time as solid lines using the left--hand axis scale. $M(\rho_{\rm crit})$ in the control simulations (blue solid lines) varies by moderate factors of a few over the simulation durations. In the feedback runs, there is generally more dense gas than in the companion control simulation by factors of a few, and again with variations over time by factors of a few.\\
\indent Star formation rates (dashed lines and right hand axis scale in Figure \ref{fig:dens_sfr}) are generally higher in the control simulations by factors of up to three compared to the corresponding feedback calculation. In both sets of calculations, the star formation rates vary non--monotonically with time by factors of approximately two.\\
\indent Since stars form from dense gas, $M(\rho_{\rm crit})/{\rm SFR}$ can be thought of as the depletion time of the dense gas due to star formation. If there are no forces present to resist gravity and gas exhaustion has not yet occurred, the depletion time would be expected to be close to the freefall time at $\rho_{\rm crit}$, $t_{\rm ff}(\rho_{\rm crit})\approx0.35$ Myr. There are no magnetic fields present in our calculations and only one of our simulations, Run F, comes anywhere near gas exhaustion, so we expect the control simulations to form stars at the freefall rate. In \cite{2014MNRAS.442..694D}, we showed that star formation in the control simulations indeed proceeds at close to the freefall rate defined by the freefall times of the entire clouds, whereas star formation in the feedback simulations was generally slower by factors up to two. In both sets of simulations, the star formation rates also vary with time, although also only by factors of approximately two over the timescales shown here.\\
\indent In Figure \ref{fig:deptimes}, we show the depletion times of gas with $\rho>\rho_{\rm crit}$ (defined as the instantaneous quantity of material satisfying this condition divided by the instantaneous star formation rate) in units of $t_{\rm ff}(\rho_{\rm crit})$ as functions of time in the control (blue lines), and dual--feedback (black lines) Runs I, J, UF, UP and UQ. The red dashed line indicates a depletion time of $t_{\rm ff}(\rho_{\rm crit})$. The depletion times in the control simulations are within a factor of two of this value for the virtually the whole duration of all simulations.\\
\indent The depletion times in the feedback runs in contrast are substantially longer, since the star formation rates in these models are slower and the dense gas masses are generally larger. The depletion times are almost always at least a factor of two longer than those in the control runs at the same epoch. Depletion times in the feedback runs also vary much more strongly with time, sometimes being more than an order of magnitude higher than in the corresponding control calculation, although the variation is often strongly non--monotonic.\\
\indent Overall, the roughly constant depletion times in the control runs indicate that star formation in these simulations proceeds quite steadily at the freefall rate in the dense gas. Since $t_{\rm ff}(\rho_{\rm crit})$ is a constant, and the star formation rates also do not vary greatly, the invariance of the depletion times in the control simulations implies that the masses of gas above $\rho_{\rm crit}$, $M(\rho_{\rm crit})$ are also nearly constant.\\
\indent In general, the depletion times in the feedback simulations are longer than $t_{\rm ff}(\rho_{\rm crit})$, sometimes by factors approaching ten, indicating a decoupling of the star formation rate from the freefall time in the dense gas. There are two extremal possible explanations for this; star formation at densities at or above $\rho_{\rm crit}$ (i) proceeds efficiently but more slowly (ii) proceeds at the same rate but less efficiently. In reality, the explanation may be a combination of these factors.\\
\begin{figure*}
     \centering
   \subfloat[Run I]{\includegraphics[width=0.19\textwidth]{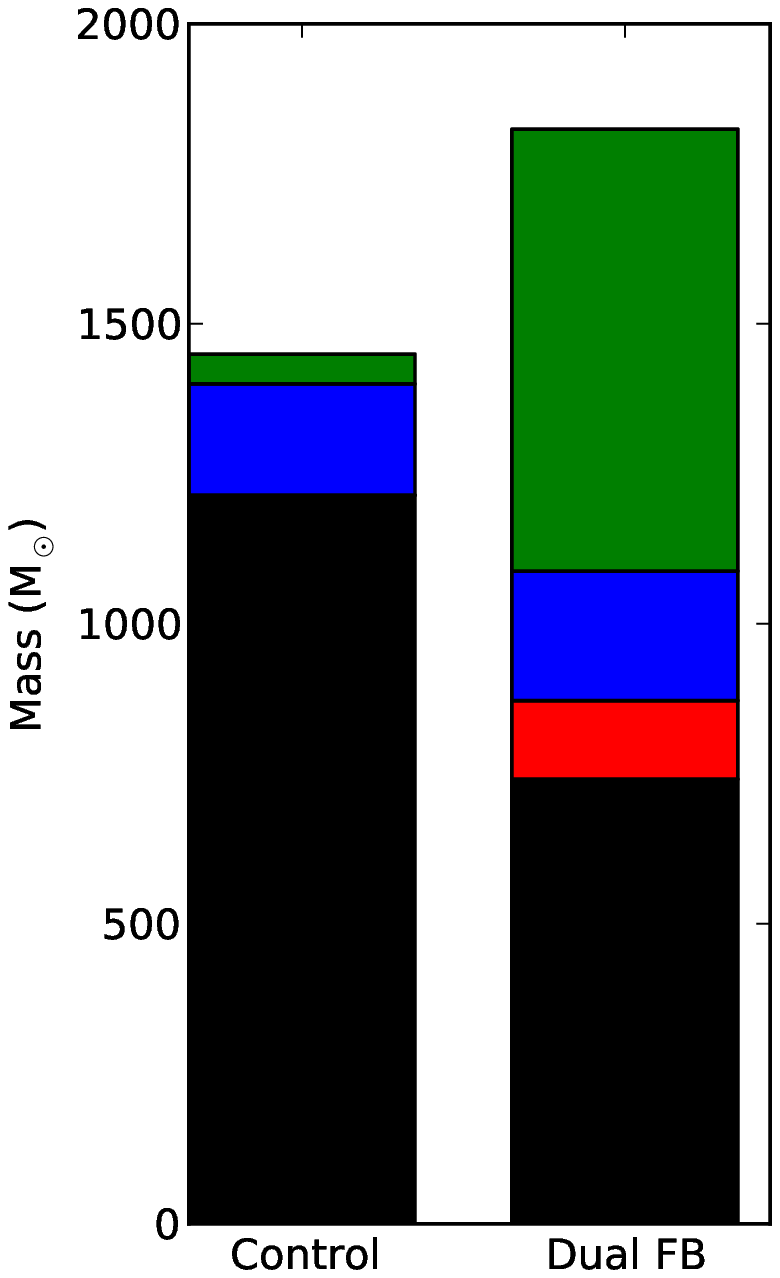}}
   \hspace{-0.05in}
      \subfloat[Run J]{\includegraphics[width=0.19\textwidth]{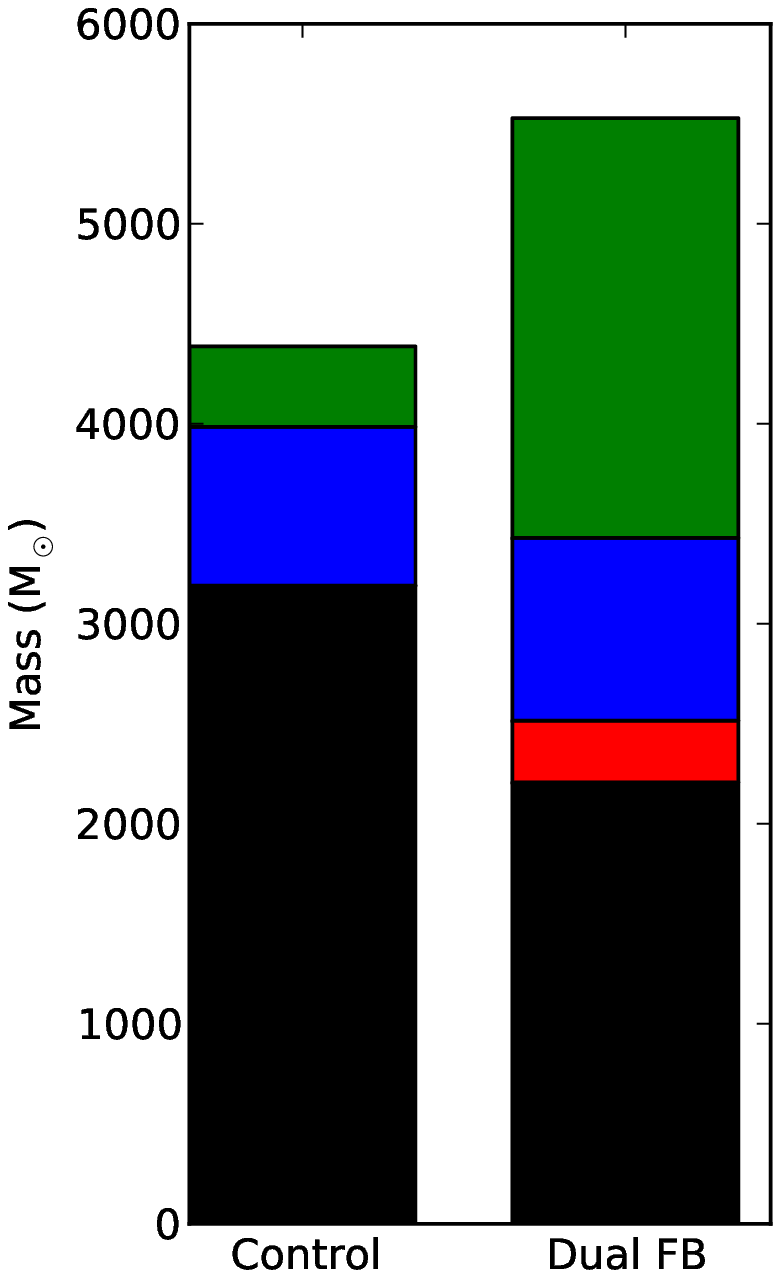}}
         \hspace{-0.05in}
      \subfloat[Run UF]{\includegraphics[width=0.19\textwidth]{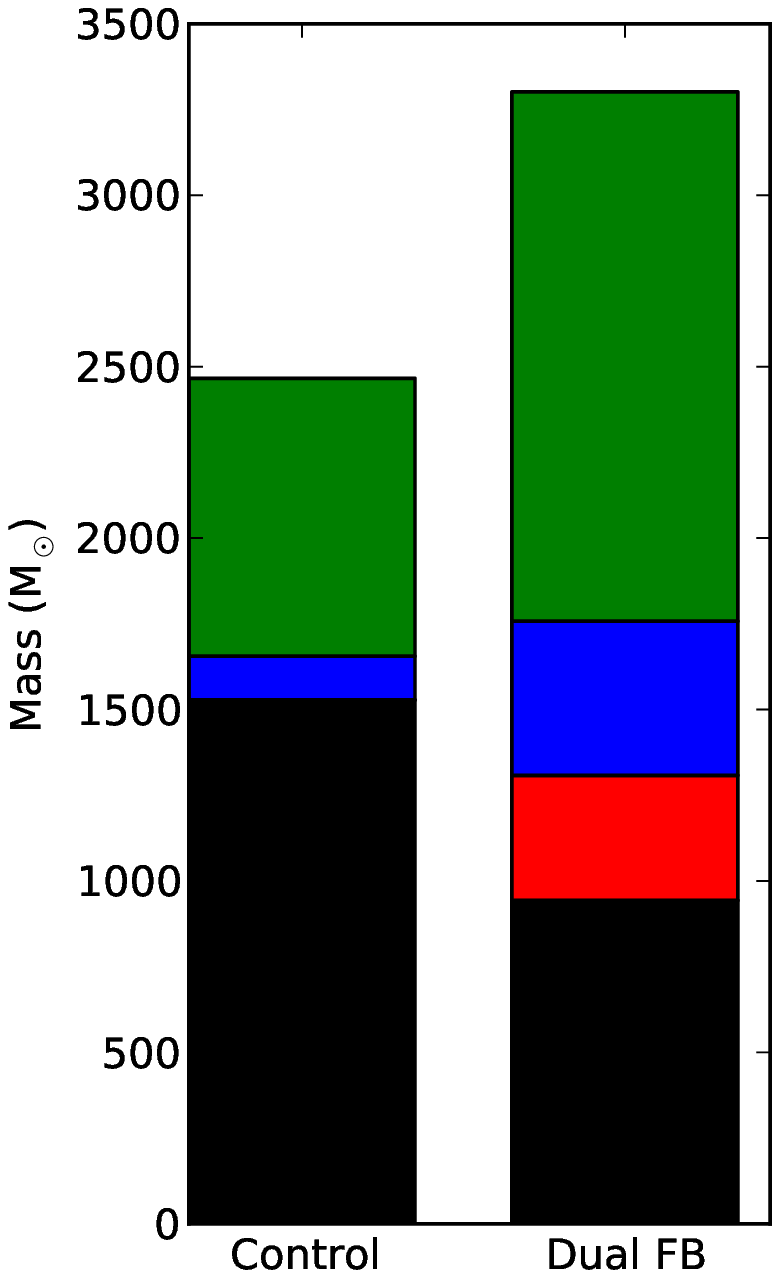}}
   \hspace{-0.05in}
      \subfloat[Run UP]{\includegraphics[width=0.19\textwidth]{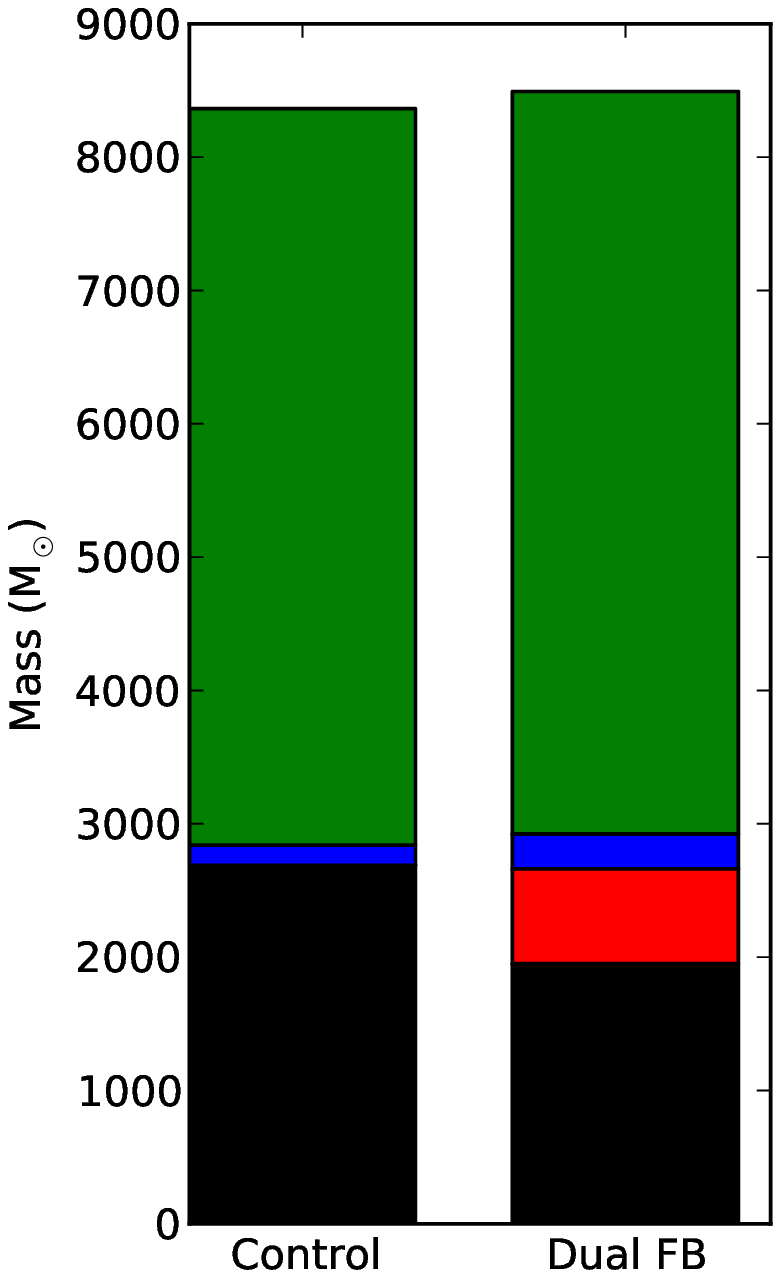}}
   \hspace{-0.05in}
      \subfloat[Run UQ]{\includegraphics[width=0.19\textwidth]{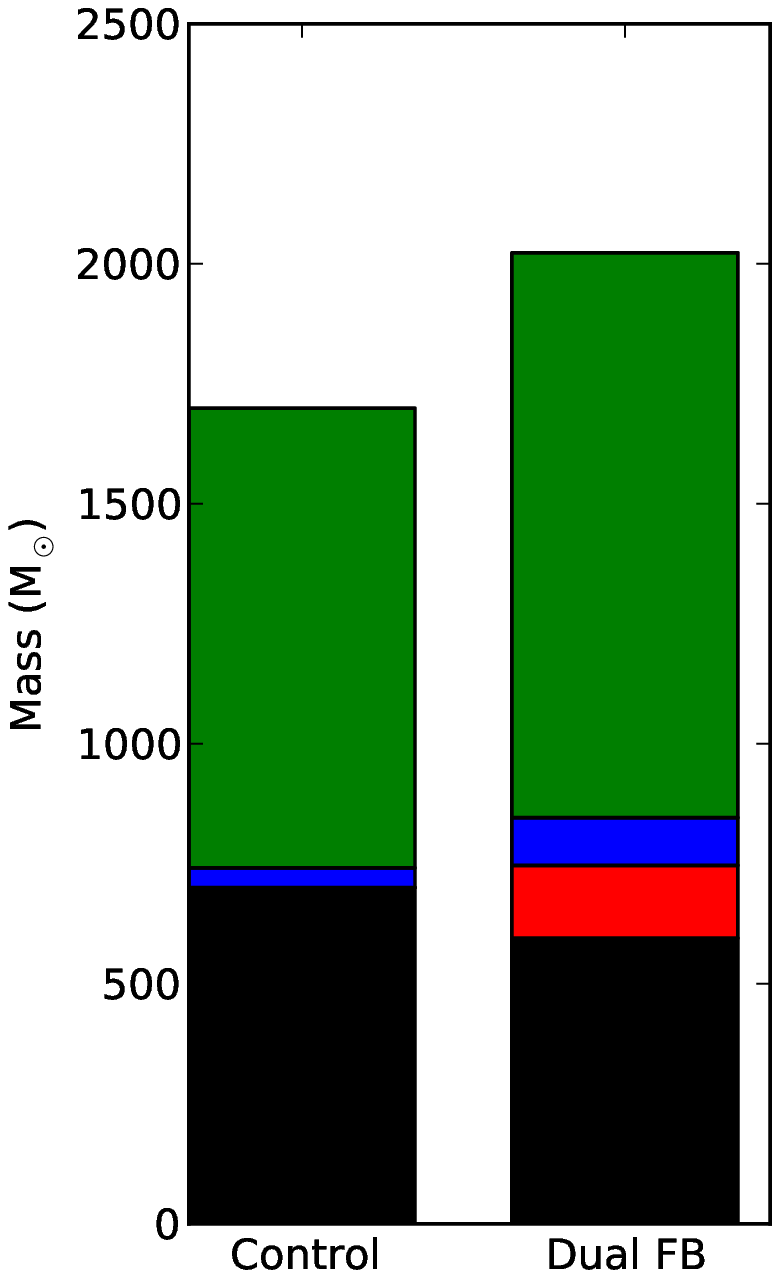}}
\caption{Barcharts recording the final fate of all gas whose density comes to exceed $\rho_{\rm crit}$ in Runs I, J, UF, UP and UQ simulations at the ends of each calculation. Lefthand columns are from simulations, righthand columns from dual--feedback calculations. Black: involved in star formation. Red: ionised. Blue: acquires higher density. Green: acquires lower density.}
\label{fig:dens_fate}
\end{figure*}
\subsection{Production of dense gas}
\indent We investigate the issue of the long dense gas depletion times in the feedback simulations by first examining how much dense gas the simulations actually produce. We identify, in all the above simulations, every SPH particle whose density \emph{ever} exceeds $\rho_{\rm crit}$, and determining the status of these particles at the end of each run. There are either three or four possibilities, depending on the calculation: (a) the material has been involved in star formation; (b) the material has been ionised; (c) the material is neutral and denser than $\rho_{\rm crit}$; (d) the material is neutral and less dense than $\rho_{\rm crit}$. In Figure \ref{fig:dens_fate} we plot bar charts for showing the results of this analysis, with the four categories of gas being shown as black, red, blue and green respectively.\\
\indent Runs I and J are globally bound clouds whose radii and therefore mean densities change little when undisturbed by feedback. In the control simulations of these clouds, the majority of gas whose density at some point in the simulations comes to exceed $\rho_{\rm crit}$ is involved in star formation by the ends of the simulations. A small amount is still in gaseous form and denser than $\rho_{\rm crit}$, and an even smaller amount has fallen to lower densities. In the feedback runs, it is evident that more dense gas is produced in total, but that less of it is eventually involved in star formation. However, the difference is not due to large quantities of dense gas being subsequently ionised, or to gas remaining at high densities but not forming stars. Instead, large quantities of gas are elevated over the density threshold, but later fall back below it, while remaining neutral.\\
\indent Runs UF, UP and UQ, are globally unbound and thus expand substantially whether feedback is acting or not. Consequently in the control simulations of these clouds, a substantial fraction (the majority in Runs UP and UQ in fact) of the gas whose density is raised above $\rho_{\rm crit}$ by turbulence (or gravity) subsequently re--expands and is never involved in star formation. We reiterate that this is nothing to do with feedback and is just a result of the initial supervirial states of these clouds. In the simulations in which feedback \emph{is} active, the fraction of dense gas involved in star formation again drops and the total quantities of dense gas produced again increase, although only marginally in the case of Run UP. The fraction of dense gas which is ionised is once more very modest, although the ionised gas is the chief difference between the control and dual--feedback Runs UP. In Runs UF and UQ, feedback results in more dense gas being produced but much of it falls back below the density threshold without being ionised.\\
\indent These plots support several general conclusions:\\
\indent (i) The total quantity of sometime--dense gas in dual--feedback simulations is larger than in the control simulations. Feedback aids these clouds in generating dense gas.\\
\indent (ii) The fractions of dense gas which finishes simulations with densities lower than $\rho_{\rm crit}$ is always greater in the feedback calculations than in the corresponding control run.\\
\indent (iii) The quantities of dense gas which are ionised in the dual--feedback calculations are in all cases small fractions of the totals. Dense gas in these simulations is therefore mostly not prevented from forming stars by being ionised.\\
\indent (iv) The final star formation efficiencies are always lower in the dual--feedback simulations. Feedback restrains the clouds from forming stars.\\
\indent The obvious question arising from these observations is, if the feedback--affected clouds are better at forming dense gas, why are they worse at making stars?\\
\begin{figure*}
     \centering
   \subfloat[Control, t$_{\rm ion}$]{\includegraphics[width=0.33\textwidth]{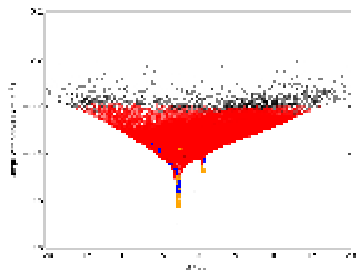}}
   \hspace{-0.1in}
      \subfloat[Control, t$_{\rm final}$]{\includegraphics[width=0.33\textwidth]{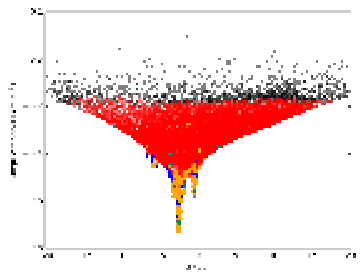}}
   \hspace{-0.1in}
      \subfloat[Dual--feedback, t$_{\rm final}$]{\includegraphics[width=0.33\textwidth]{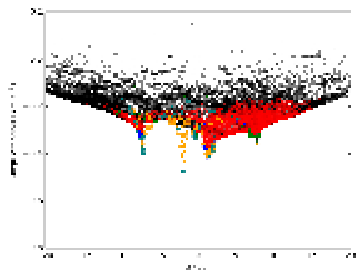}}
\caption{The potential wells of the control simulation at t$_{\rm ion}$ (left) and t$_{\rm final}$ (centre) and the dual--feedback simulation at t$_{\rm final}$ (right) in Run I. Black dots are particles that are unbound in the cluster centre--of--mass frame and red dots are unbound particles. Bound particles with densities above $\rho_{\rm crit}$ are shown in blue, and unbound particles exceeding this density are shown in green. Orange and teal triangles represent bound and unbound sink particles respectively. Unbound, in all cases, means having positive total energy in the system centre--of--mass frame. Note that only one tenth of all particles in each class are plotted.}
\label{fig:runi_wells}
\end{figure*}
\begin{figure*}
     \centering
   \subfloat[Control, t$_{\rm ion}$]{\includegraphics[width=0.32\textwidth]{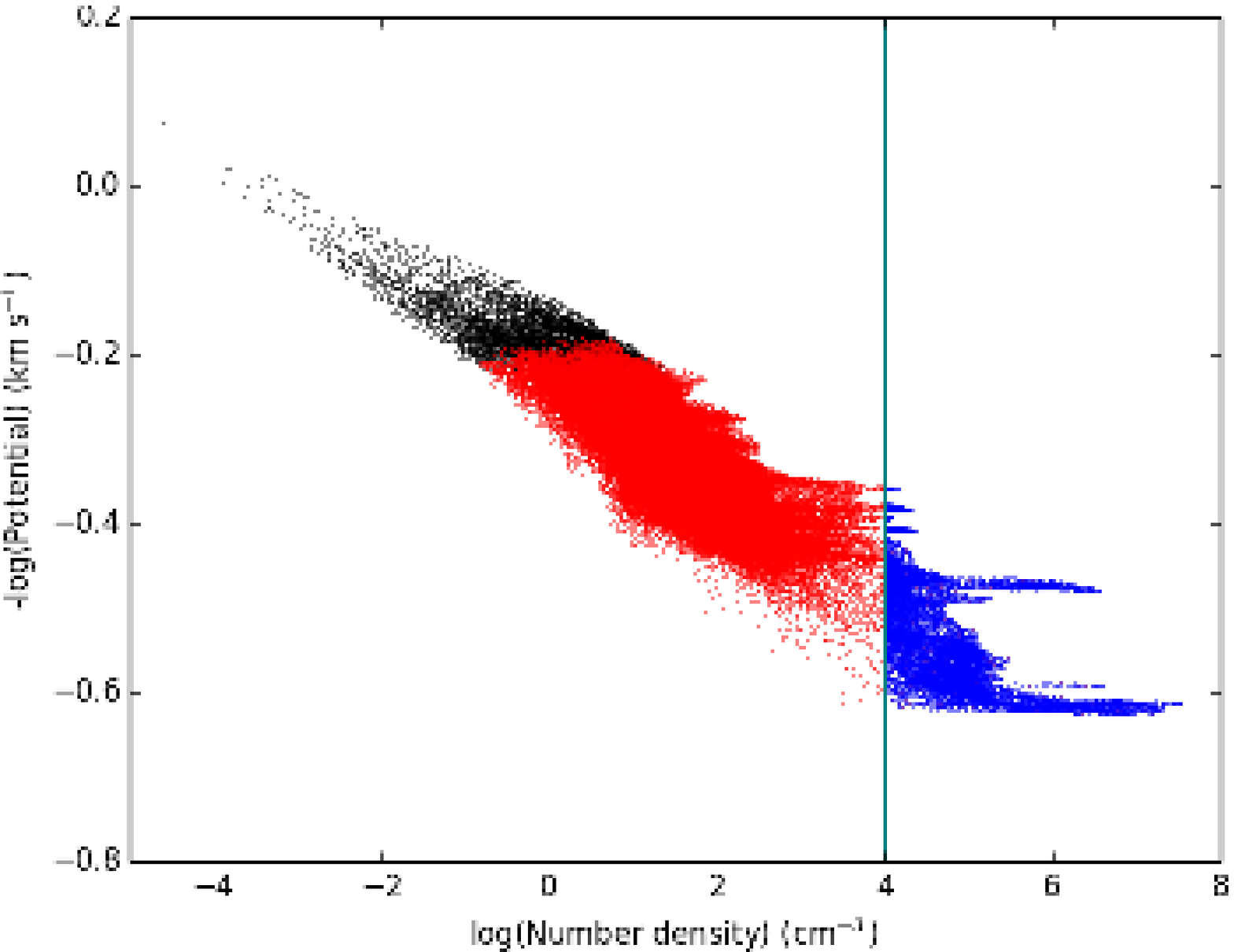}}
   \hspace{0.1in}
      \subfloat[Control, t$_{\rm final}$]{\includegraphics[width=0.32\textwidth]{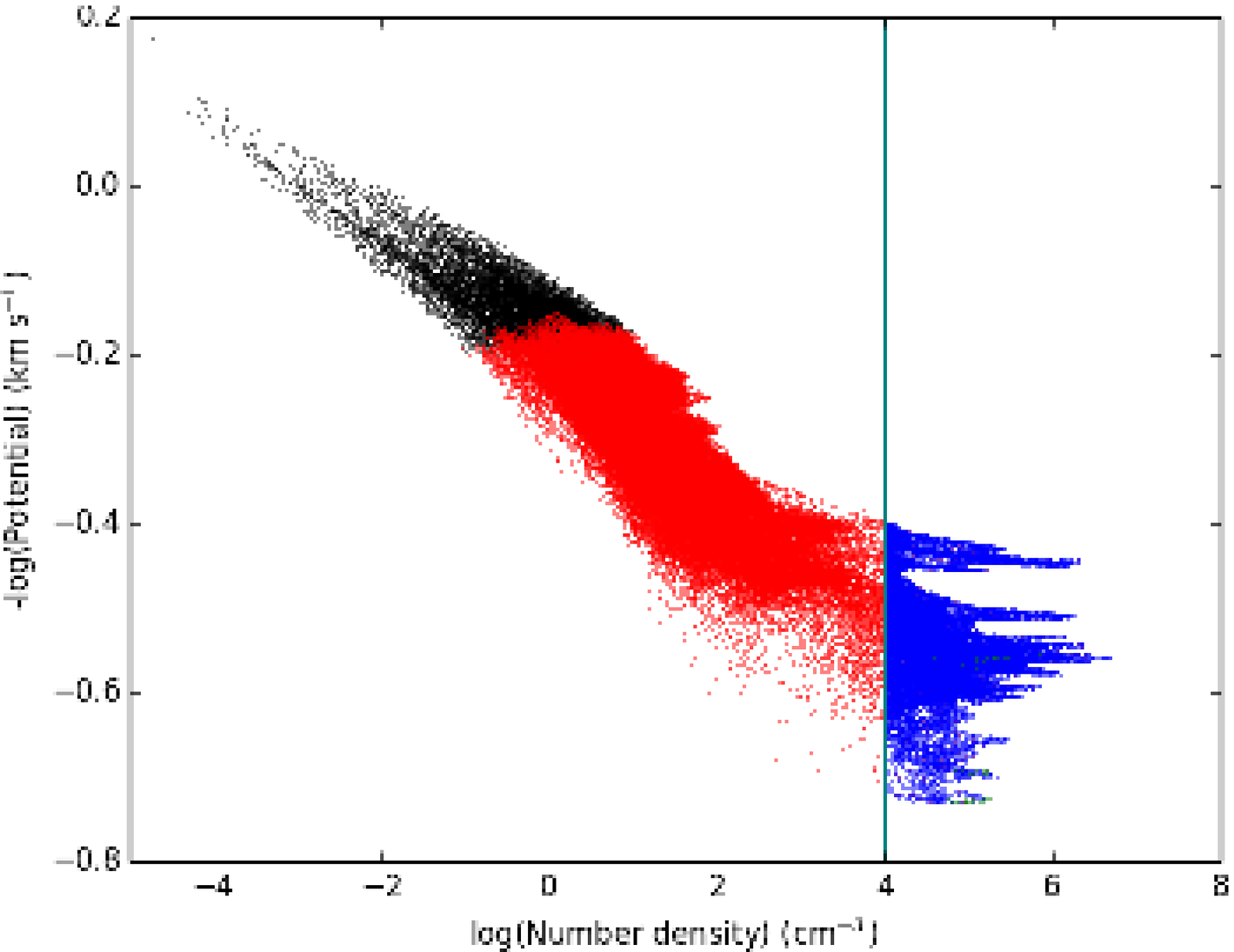}}
   \hspace{0.1in}
      \subfloat[Dual--feedback, t$_{\rm final}$]{\includegraphics[width=0.32\textwidth]{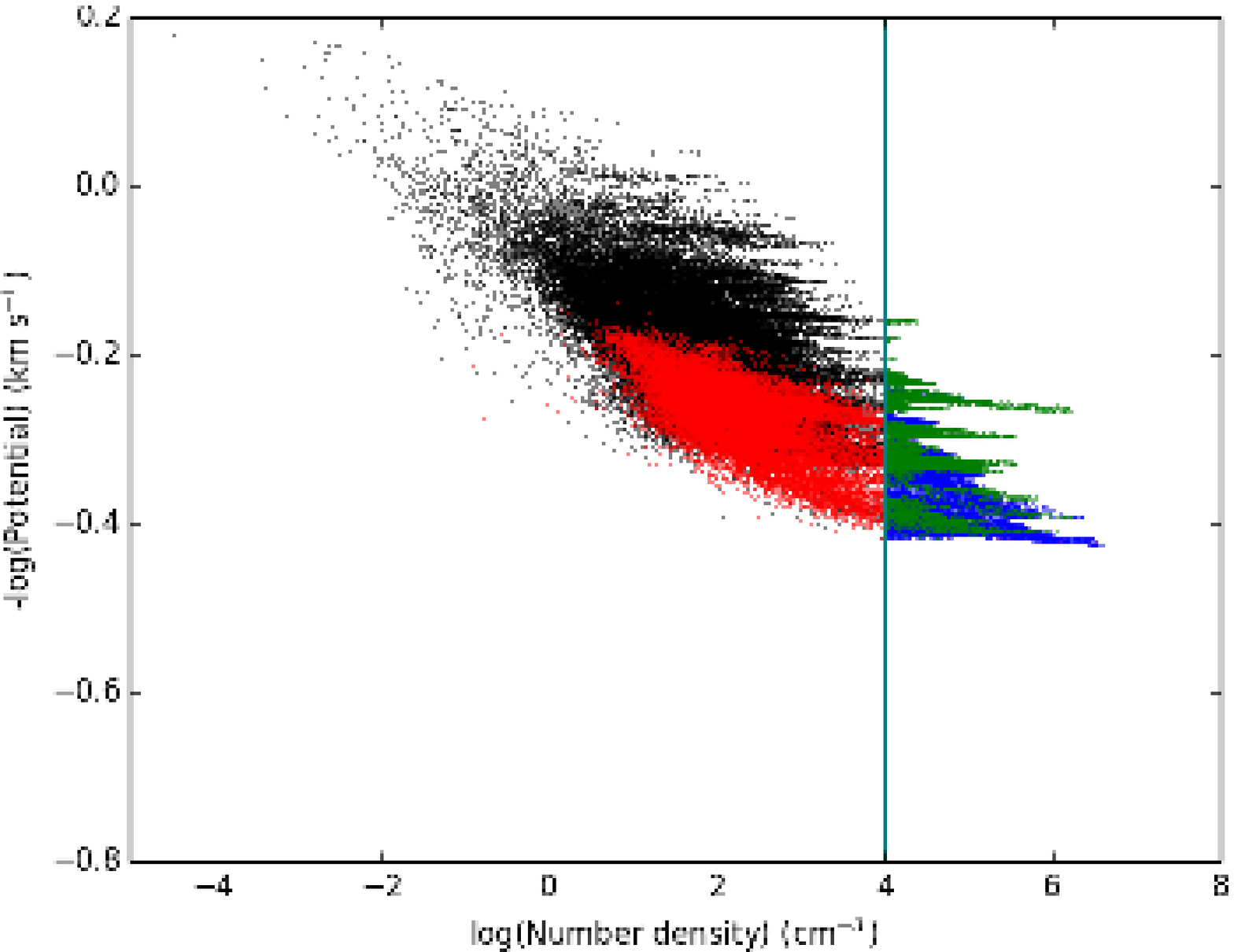}}
\caption{Gravitational potential against density at t$_{\rm ion}$ (left) and t$_{\rm final}$ (centre) in the control run, and in the dual--feedback simulation at t$_{\rm final}$ (right) in Run I. Black dots are particles that are unbound in the cluster centre--of--mass frame and red dots are unbound particles. Bound particles with densities above $\rho_{\rm crit}$ are shown in blue, and unbound particles exceeding this density are shown in green. Unbound, in all cases, means having positive total energy in the system centre--of--mass frame. The vertical teal line is $\rho_{\rm crit}$. Note that only one tenth of all particles in each class are plotted.}
\label{fig:runi_phi_dens}
\end{figure*}
\subsection{Conversion of dense gas to stars}
The answer to the question of why the feedback calculations are not efficient at converting their gas to stats lies in the fact that dense gas is a \emph{necessary but not sufficient} condition for the production of stars. The gas must also be gravitationally bound and collapsing. In Figure \ref{fig:runi_wells}, we show the potential wells from the pre--feedback (left) and final (centre) states of the control Run I simulations, together with the final state of the dual--feedback simulation (right). Equivalent plots from runs J, UF, UP and UQ are very similar and are not shown to save space. Gas particles are shown as small dots. Only non--ionised gas particles are shown and, for reasons of clarity, only every tenth particle is plotted. Bound particles whose densities exceed $\rho_{\rm crit}$ are shown in blue, unbound particles exceeding this density in green, and other unbound and bound particles as black or red respectively. Sink particles are shown as triangles, with orange indicating bound sinks and teal unbound sinks. Several points are immediately apparent.\\
\indent (i) The potential well in the control simulation becomes deeper and broader with time as more gas falls into it and more stars are formed. The stars generally congregate in the deepest part of the well. In the final states of the control and dual--feedback simulations, there are a few unbound sink particles deep within the respective potential wells. These are low--mass sinks that have become unbound by dynamical interaction with more massive partners.\\
\indent (ii) Unbound gas particles in the control simulation are rather uniformly--distributed at values of the logarithm of the potential (expressed in kilometres per second) close to zero. In the ionised run, by contrast, unbound gas particles exist at substantially greater depths in the potential well, due to local acceleration of gas to velocities higher than the cloud escape velocity.\\
\indent (iii) Gas whose density is larger than $\rho_{\rm crit}$ resides almost exclusively in the deepest troughs in the potentials. This is particularly obvious in the control run.\\
\indent (iv) In the control simulation, the densest gas is coincident with the densest groups of stars in the deepest potential troughs. However, in the dual feedback simulation, the deepest potential well is devoid of gas and contains only stars. This well is substantially shallower than its counterpart in the final state of the control simulation. The dual--feedback simulation instead possesses several smaller shallower potential wells containing mixtures of gas and stars, and not all of the material in these wells is bound in the cloud centre--of--mass frame.\\
\indent Our interpretation of these observations is as follows. The cloud's primary potential well in the control run is the main engine for the production of dense gas and for the conversion of gas to stars. The depth and dominance of this well ensures that gas continually flows into it and that this gas is bound and has little option but to form stars. In the feedback run, the reversal of the accretion flows feeding the central cluster and the expansion of the cluster itself has prevented the potential well acquiring the depth it achieves in the control simulation. Additionally, the deepest potential well in the feedback simulation has been completely emptied of gas, so it is not able to contribute to star formation. Compression of gas on the peripheries of the cleared--out bubbles has instead produced several less massive and shallower potential troughs. There is thus no large and efficient central star factory in the dual--feedback simulation and none of the gas has access to a potential as deep as that in the control simulations.\\
\indent The expansion of the HII regions/wind bubbles into the outer regions of the clouds in the dual--feedback simulations does create additional gas whose densities exceed $\rho_{\rm crit}$ for some time, but this gas never becomes gravitationally bound. We illustrate this in Figure \ref{fig:runi_phi_dens}, which depicts the potential of the gas as a function of density for the control run at the time of the initiation of feedback (left panel) and the end of the simulation (centre panel), compared to the end of the dual--feedback run (right panel). Black and red particles are unbound and bound particles respectively with densities below $\rho_{\rm crit}$, while green and blue particles are unbound and bound particles with densities above $\rho_{\rm crit}$.\\
\indent In the control simulations at both epochs, all gas above $\rho_{\rm crit}$ is bound in the centre--of--mass frame, and thus highly likely to form stars. In the dual--feedback run, by contrast, there are large quantities of unbound gas whose density exceeds $\rho_{\rm crit}$. Although this gas is unbound in the frame of the cloud centre of mass, this does not necessarily imply that none of it will become locally self--bound. However, as discussed in \cite{2013MNRAS.436.3430D}, because of the steep mean radial density profiles ($\sim r^{-2}$) of the clouds, the surface density of swept--up material declines with time and radius and this gas becomes more gravitationally stable as the simulation progresses. We find that large quantities of gas which become denser than $\rho_{\rm crit}$ fail to become bound and eventually fall back below this density as they are driven to larger radii. This argues against a strict density threshold for star formation in these calculations.\\
\begin{figure}
\includegraphics[width=0.48\textwidth]{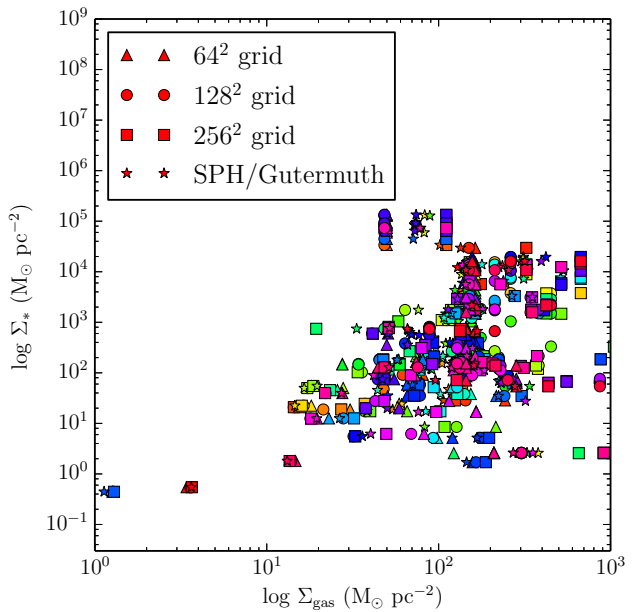}
\caption{ Stellar mass surface density plotted against gas mass surface density for the final state of the control Run I simulation, at the location of each sink particle. Four points in a randomly--chosen colour are shown for each sink particle. In all cases, the \emph{stellar} mass surface density is computed using the method of Gutermuth et al. 2011, with N=10. For the star symbols, the \emph{gas} mass surface density has been computed by evaluating an SPH surface density sum at the locations of each sink particle. For the triangle, circle and square symbols, a uniform grid of pixels has been placed on a 15$\times$15pc box and the surface density computed by performing an SPH surface density sum at location of the pixel centres. Each sink particle has then been assigned the surface density of the pixel in which it s projected. Triangles represent results for a 64$^{2}$ grid, circles a 128$^{2}$ grid and squares a 256$^{2}$ grid.}
\label{fig:sigma_npix}
\end{figure}
\subsection{Stellar and star formation rate surface densities}
\begin{figure*}
     \centering
   \subfloat[Run I]{\includegraphics[width=0.33\textwidth]{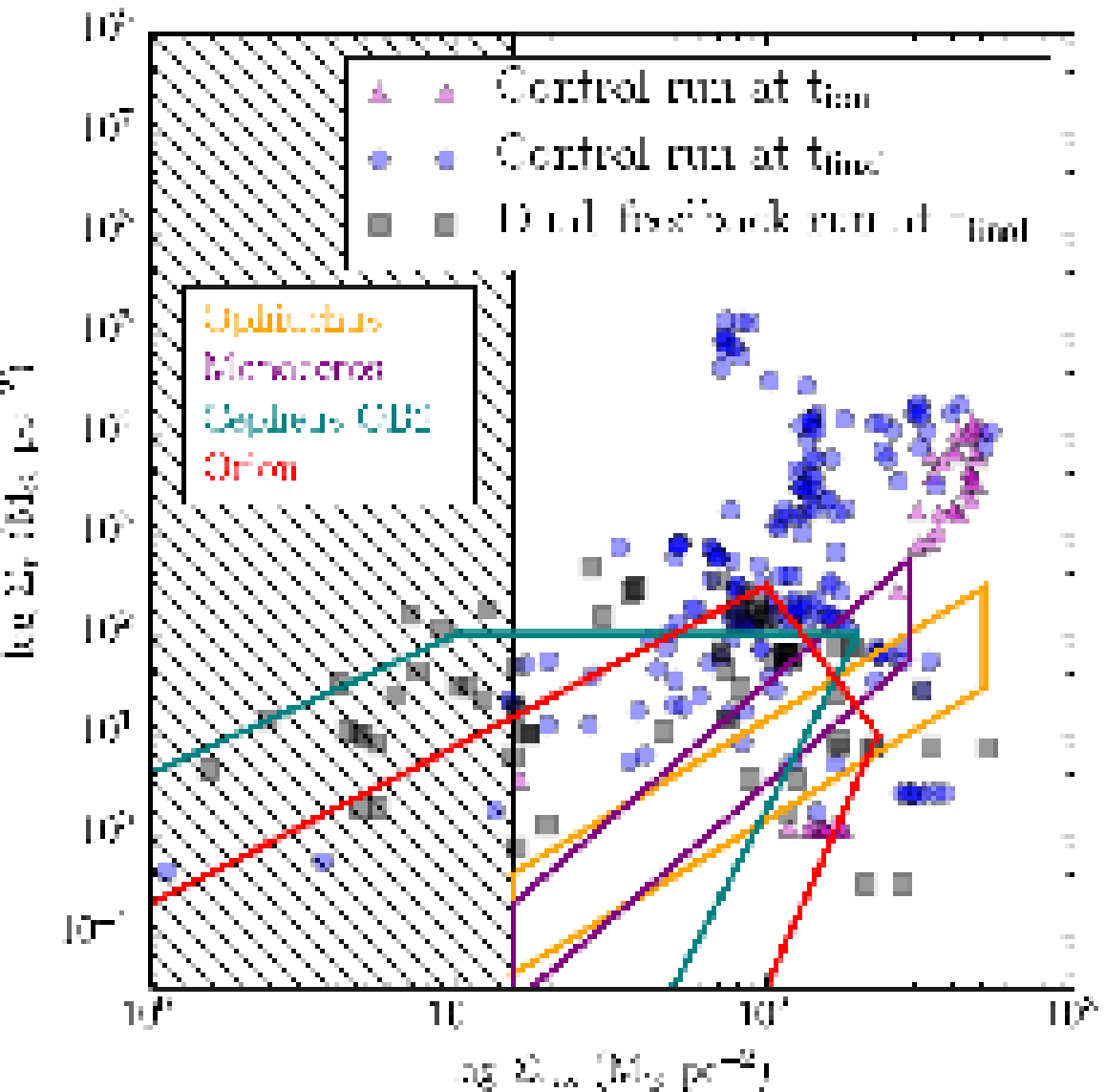}}
   \hspace{-0.05in}
      \subfloat[Run UF]{\includegraphics[width=0.33\textwidth]{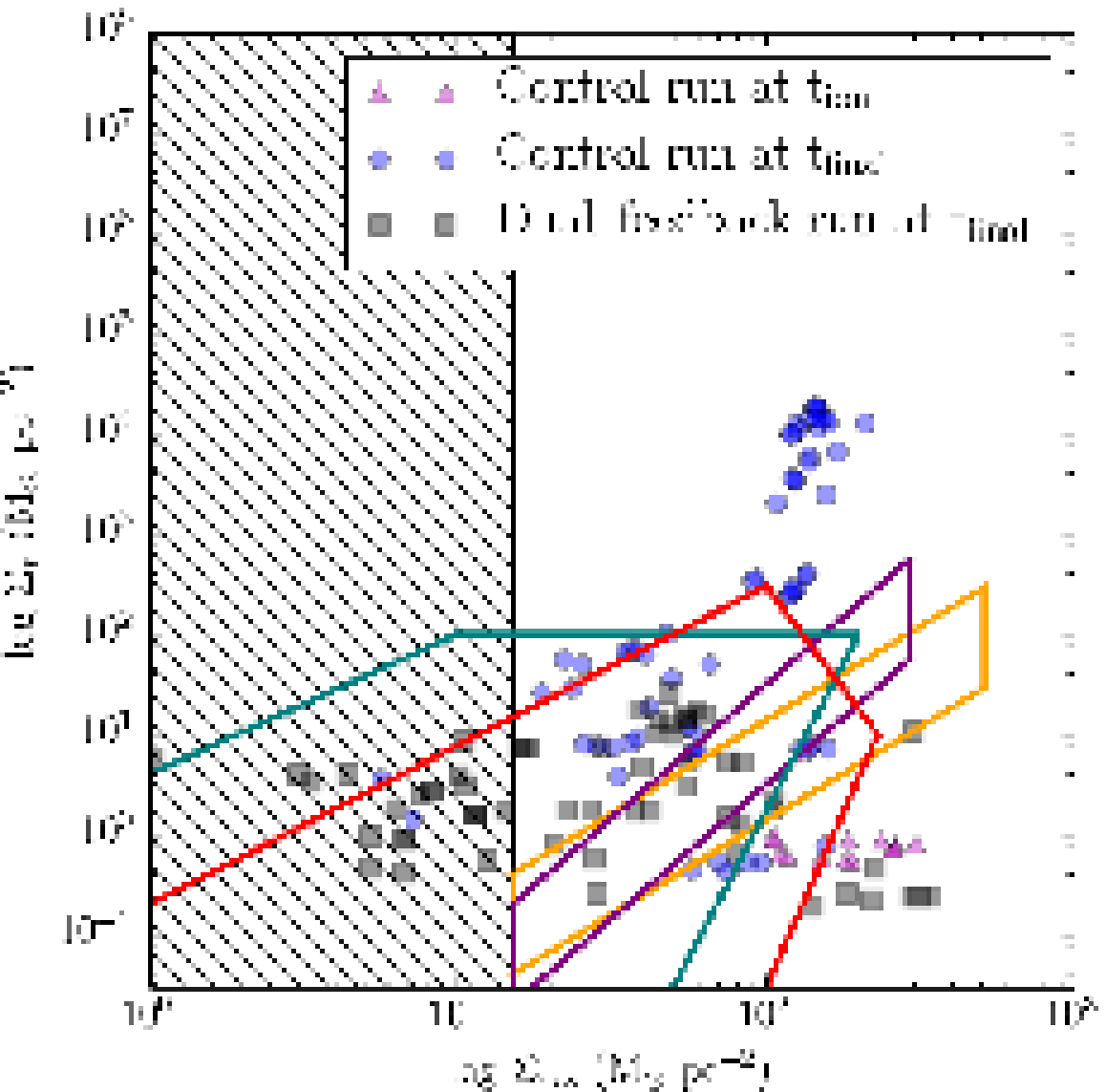}}
         \hspace{-0.05in}
      \subfloat[Run E]{\includegraphics[width=0.33\textwidth]{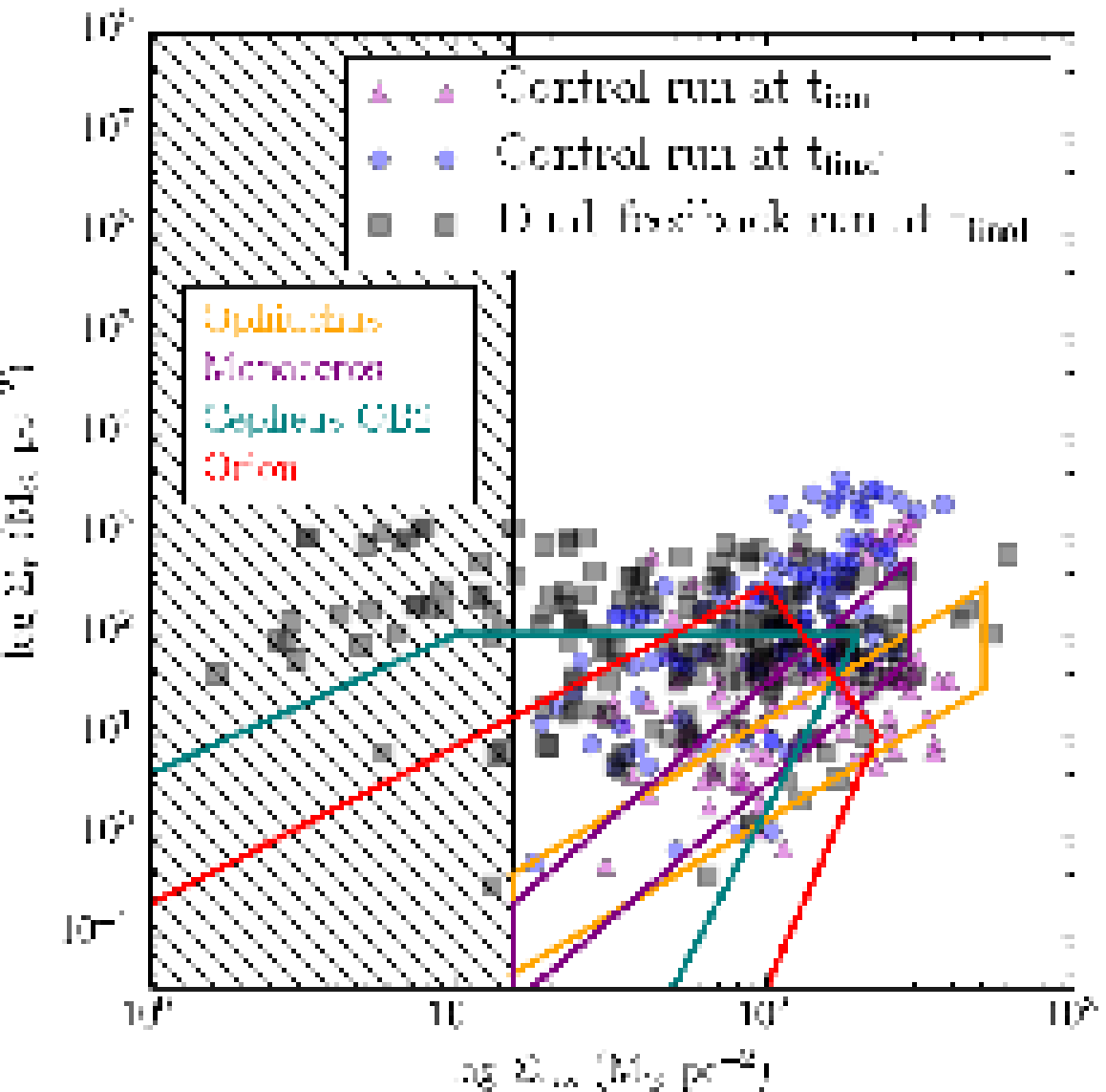}}
\caption{Stellar mass surface densities versus gas mass surface densities at the locations of all sink particles for Runs I (10$^{4}$M$_{\odot}$ bound cloud, left panel), UF (3$\times$10$^{4}$M$_{\odot}$ unbound cloud, centre panel) and E (10$^{5}$M$_{\odot}$ bound cloud, right panel). Red triangles are from times when feedback is initiated, blue circles are from the end of the control runs, black squares are from the end of the dual--feedback runs. Hatched areas have $A_{\rm V}<1$, where observational results are likely to become unreliable. The polygons are from Gutermuth et al. 2009's results; orange is Ophiuchus, purple is Mon R2, teal is Ceph OB2 and red is Orion.}
\label{fig:smass_sigma}
\end{figure*}

\begin{figure*}
     \centering
   \subfloat[Run I]{\includegraphics[width=0.33\textwidth]{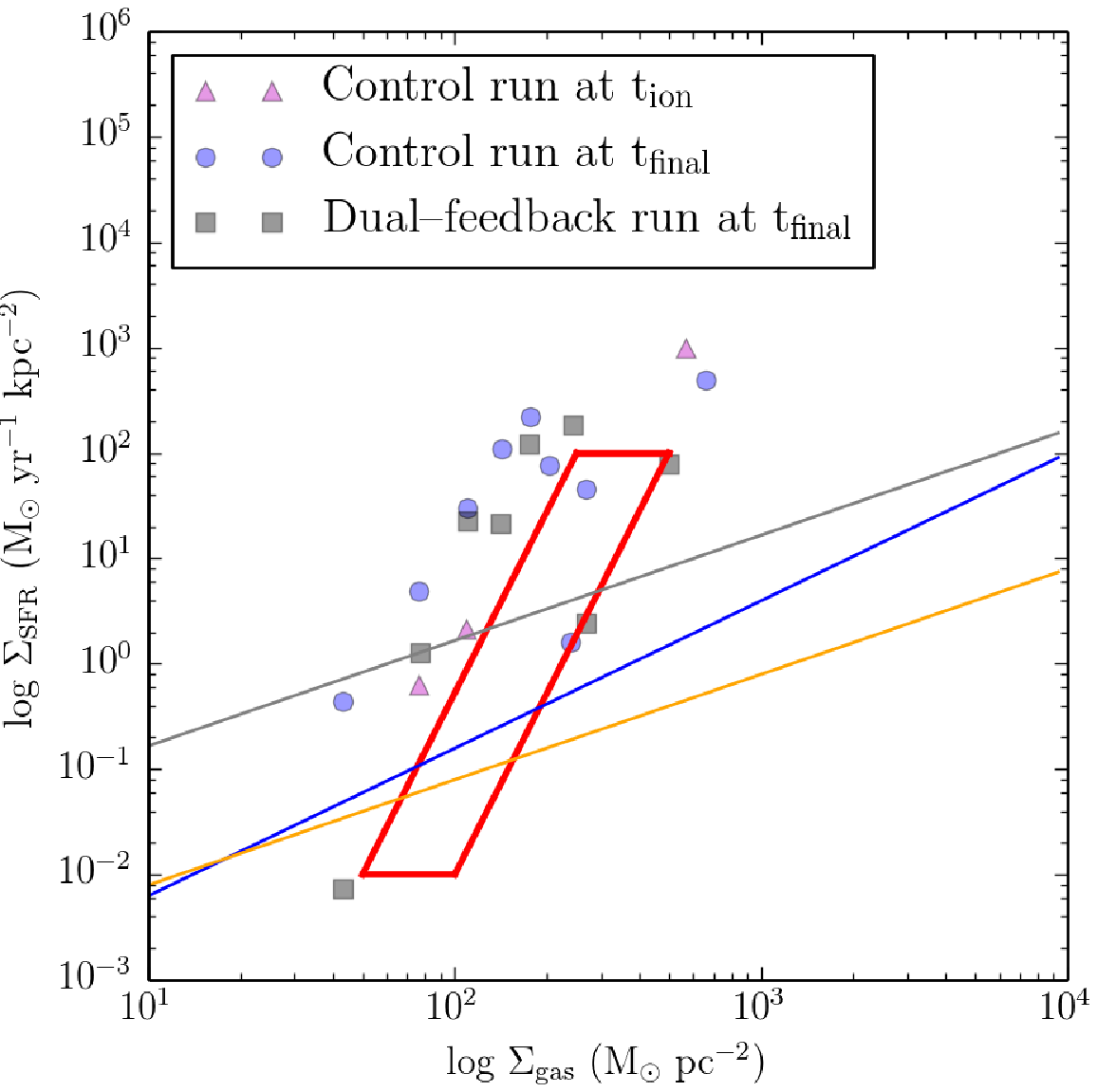}}
   \hspace{-0.05in}
      \subfloat[Run UF]{\includegraphics[width=0.33\textwidth]{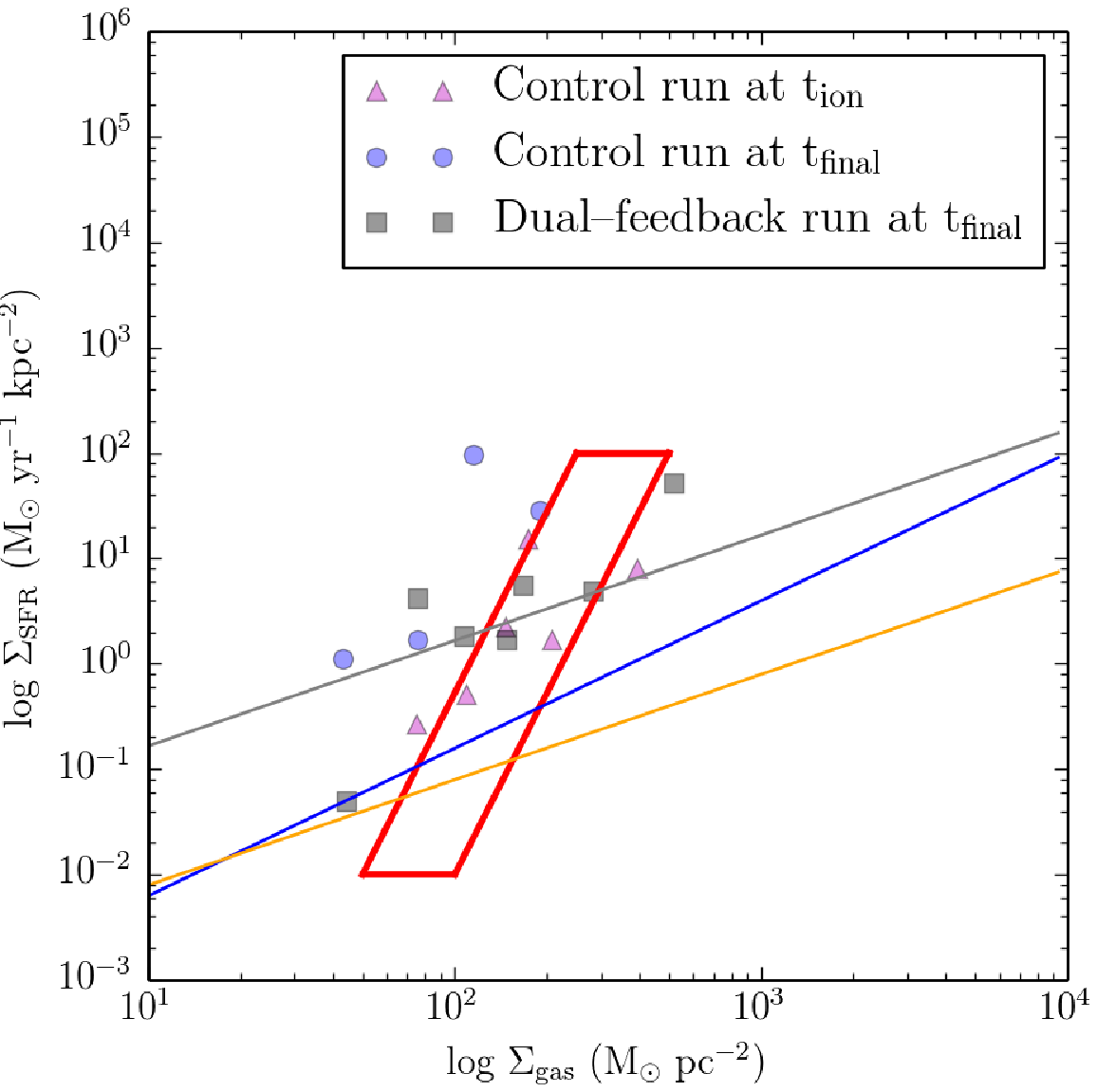}}
         \hspace{-0.05in}
      \subfloat[Run E]{\includegraphics[width=0.33\textwidth]{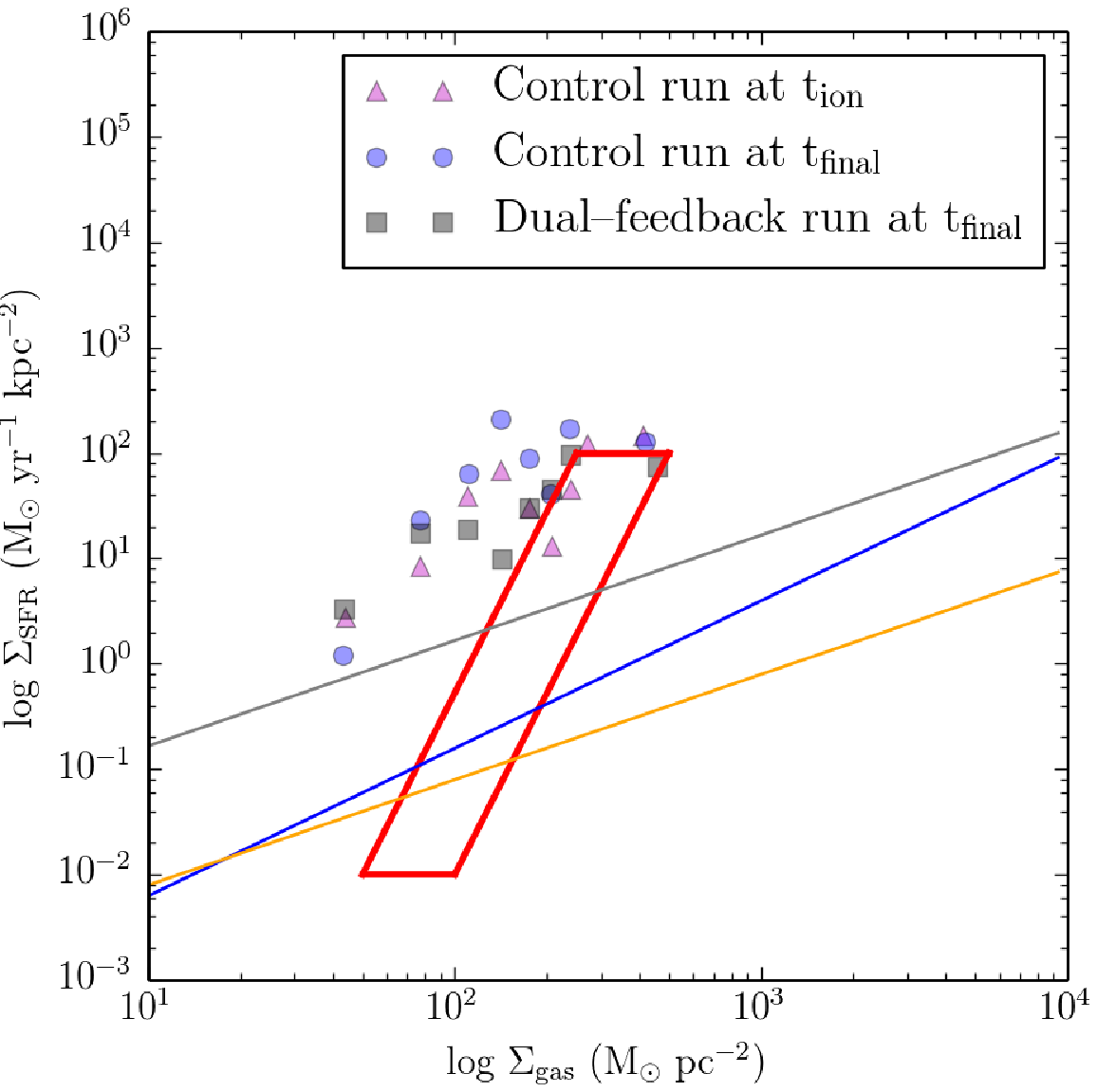}}
\caption{Star formation rate surface densities versus gas mass surface densities in Runs I (10$^{4}$M$_{\odot}$ bound cloud, left panel), UF (3$\times$10$^{4}$M$_{\odot}$ unbound cloud, centre panel) and E (10$^{5}$M$_{\odot}$ bound cloud, right panel). Pink triangles are from times when feedback is initiated, blue circles are from the end of the control runs, black squares are from the end of the dual--feedback runs. The black line shows the relation form Wu et al., 2005, the blue line is the Kennicutt et al., 1998 relation and the orange line is from Bigiel et al., 2008. The red region delineates approximately the region covered by the c2d and Gould Belt clouds as reported by Heiderman et al. (see their Figure 3).}
\label{fig:sfr_sigma}
\end{figure*}

\indent In \cite{2014MNRAS.442..694D} and in the preceding subsections, we discussed chiefly the global properties of the model clouds, such as their final star formation efficiencies and mean star formation rates. Here we examine the relations between resolved quantities, namely the stellar mass surface density or star formation rate surface density, and the gas mass surface density.\\
\indent \cite{2011ApJ...739...84G} used resolved observations of eight nearby molecular clouds to examine the correlation between the stellar mass surface density and gas mass surface density. Gas surface densities are computed from near--IR dust extinction mapping, and the cloud total masses are computed from summing the mass at extinctions above detection thresholds peculiar to each cloud. The most massive system examined was Orion at 33 200 M$_{\odot}$, and the least massive was Serpens at 2 590 M$_{\odot}$, with a mean of $\approx$14 500 M$_{\odot}$, comparable to our low mass runs I, J, UP, UQ and UF. The A$_{\rm V}$ thresholds used for these determinations varied from -1.0 (for Orion) to 3.7 (for Serpens), with a mean of $\approx 0.7$, or $\approx 0.3$ if Serpens is excluded. These figures illustrate the problem of computing the mass of a GMC observationally. \cite{2011ApJ...739...84G} computed stellar surface densities by drawing a circle of radius $R_{\rm N}$ around each star cutting through the position of the $N$th nearest neighbouring star, and computing the surface density as $(N-1)/\pi R_{\rm N}^{2}$, using $N=10$.\\
\indent We replicate these techniques as closely as possible. We use the same method as \cite{2011ApJ...739...84G} to estimate the stellar density at the sink particle locations. We also note that typically 80--90$\%$ of the mass of our model clouds exists at $A_{\rm V}>$0.5, so that their actual masses are close to what would be measured by extinction mapping with the typical thresholds used by \cite{2011ApJ...739...84G}.\\
\indent However, it is not immediately obvious how to best compute the gas mass surface densities, so we employ two techniques which are compared in Figure \ref{fig:sigma_npix}, using the control Run I simulation as a test. The results plotted in this figure are from the $z$--axis projection of the simulation, but we confirmed that observing along the $x$-- or $y$--axes did not produce substantially different results. We first measured the gas surface density at the position of every sink particle by integrating through the smoothing kernels of every gas particle which overlaps in projection the position of the sink, to compute the projected mass as viewed along the $z$--axis. The same technique is used over a uniform grid to generate the column density images in our other papers. This technique is in some sense scale--free, since no grid is imposed on the density field. Alternatively, we placed uniform grids of 64$^{2}$, 128$^{^2}$ and 256$^{2}$ on the clouds and computed the column--density in each pixel by performing an SPH column--density sum at the location of each pixel centre. We then assign column--densities to the sink particles using the value of the pixel in which the sink is projected to lie. Figure \ref{fig:sigma_npix} shows that these four different estimations of the gas surface density agree tolerably well over the range of grid resolutions tried. The 64$^{2}$ grid is not able to capture the highest column densities, greater than $\sim3\times10^{2}$M$_{\odot}$pc$^{-2}$, but otherwise the forms of the plots are rather similar. Since we are only going to use them for making qualitative judgements, we consider either of the techniques used adequate, and we adopt the first one, in which the column densities are computed at the sink particle positions.\\
\indent In Figure \ref{fig:smass_sigma}, we plot stellar surface densities against gas surface densities for Runs I, UF and E, comparing the results at the epoch when feedback is enabled (red triangles) to those from the ends of the control runs (blue circles) and the ends of the dual--feedback runs (black squares).\\
\indent In all three pairs of simulations, the stellar surface densities grow in the control simulations relative to the epochs when feedback is enabled, by almost five orders of magnitude in the case of Run UF. The maximum gas surface densities, by contrast, are generally somewhat lower at the ends of the control simulations, reflecting consumption of gas. The dual--feedback simulations all exhibit lower maximum stellar surface densities, due to the general result of feedback that the stellar systems have lower volume-- and surface--densities, due to weakening of the local potential and decreases in the star formation efficiencies. The maximum gas surface densities are somewhat higher in the dual feedback simulations and tend not to be associated with high stellar surface densities, reflecting the relative inability of feedback to trigger star formation discussed in \cite{2013MNRAS.431.1062D}.\\
\indent We compare our results to those of \cite{2011ApJ...739...84G} (their Figure 9). Included in Figure \ref{fig:smass_sigma} are coloured polygons delineating approximately the areas of parameter space occupied by Ophiuchus (orange), Mon R2 (purple), Cepheus OB2 (teal) and Orion (red). The control simulations, particularly of Runs I and UF, resemble most in \emph{form} Ophiuchus and Mon R2, in that generally higher stellar surface densities correspond to higher gas densities. This correlation is quite tight in the case of the observations, as illustrated by the orange and purple polygons, and the relation between stellar mass and gas mass surface density follows approximately $\Sigma_{*}\propto\Sigma_{\rm gas}^{1.87}$ and $\Sigma_{*}\propto\Sigma_{\rm gas}^{2.67}$.\\
\indent The control Run I simulation initially exhibits a tight correlation between $\Sigma_{*}$ and $\Sigma_{\rm gas}$ but the relations at the ends of the control runs all show more substructure and considerably weaker correlations. The formal fitted slopes for the final states of the control runs I, UF and E are 0.99, 2.58 and 1.70 respectively, so that the slopes of the $\Sigma_{*}/\Sigma_{\rm gas}$ relations for runs UF and E are similar to those of Ophiuchus and Mon R2. However, we also observe substantially higher maximum stellar mass surface densities in the control simulations than are shown by \cite{2011ApJ...739...84G}. This likely has several causes. Resolution limitations in the observations underestimate the surface densities of the densest regions in the real clouds, but Ophiuchus, being a low--mass star--forming region is not expected to exhibit very high stellar surface densities in any case. However, probably more importantly, our control simulations where star formation is entirely unrestrained produce too many stars in configurations which are very compact, leading to very high stellar surface densities\\
\indent The feedback simulations, in contrast, produce substantially lower stellar surface densities more similar to those observed by \cite{2011ApJ...739...84G}, particularly in the case of Runs I and UF, where the maximum stellar densities are reduced by approximately two orders of magnitude relative to the control runs at the same epoch. The feedback calculations resemble more \cite{2011ApJ...739...84G}'s Cepheus OB3 and Orion data, illustrated by the teal and red polygons respectively. The simulations and observations exhibit large horizontal spreads, corresponding to regions with high stellar densities that have been partially or largely cleared of gas. Termination of accretion flows, stifling of cluster growth, and spatial separation of stars and gas in the feedback simulations produce stellar surface densities and correlations between stellar and gas mass surface densities closer to those from Gutermuth et al. in systems where feedback active.\\
\indent \cite{2010ApJ...723.1019H} instead plotted the star formation rate surface density for twenty nearby c2d and Gould's Belt clouds with masses ranging from 189 $M_{\odot}$ (Lupus IV) to 24 400 $M_{\odot}$ (Serpens--Aquila). They also use extinction mapping, complemented with $^{12}$CO and $^{13}$CO observations. They divide the clouds into regions enclosed between extinction contours and compute gas and star formation rate surface densities for each contour for each cloud. They compute the star formation rate surface densities from YSO surface densities, assuming a mean YSO mass and age.\\
\indent We also partition our clouds into contours based on extinction. We choose a minimum $A_{\rm V}$ of 4 magnitudes and a contour spacing of 4 magnitudes, with a maximum number of contours of 8. Within each region defined by two contours, we compute the mean gas surface density. Since our sink particles are not all of the same mass and we have time information available, we obtain star formation rate surface densities in each region by locating all sink particles within that region and referring back to previous dumps 0.5Myr in the past to compute the total gain in stellar mass in each region.\\
\indent We plot the results in Figure \ref{fig:sfr_sigma}, again for Runs I, UF and E, with pink triangles from times when feedback is initiated, blue circles from the end of the control runs and black squares from the end of the dual--feedback runs. As did \cite{2010ApJ...723.1019H}, we also include the star formation rate surface density versus gas surface density relations from \cite{2005ApJ...635L.173W,1998ApJ...498..541K} and \cite{2008AJ....136.2846B} to guide the eye. Note, however, that the \cite{1998ApJ...498..541K} and \cite{2008AJ....136.2846B} relations are derived at much larger size scales and thus effectively smear out star formation activity over large areas. They are thus not expected to give an accurate representation of star formation at the level of individual molecular clouds. The results from \cite{2010ApJ...723.1019H}, however, are derived at mass and scales similar to our simulated clouds or to subregions of them, so we plot red polygons which approximately delineate the area occupied by their c2d and Gould's Belt results (see their Figure 3).\\
\indent Comparing our results to the models of \cite{2005ApJ...635L.173W,1998ApJ...498..541K} and \cite{2008AJ....136.2846B}, we find that our star formation rate densities are almost always substantially higher at a given density than predicted by the \cite{1998ApJ...498..541K} or \cite{2008AJ....136.2846B} relations. Again, this is likely due in part to the fact that these relations are derived from observations at scales of $\sim100$pc and above, so that star formation is not well resolved. At gas surface densities close to 10$^{2}$M$_{\odot}$pc$^{-2}$, the agreement with \cite{2005ApJ...635L.173W} is rather better, but at higher gas surface densities, we again recover much larger star formation rate surface densities in Runs I and E, although the agreement in Run UF is rather closer.\\
\indent Our results are in fact much closer in form to those of \cite{2010ApJ...723.1019H}'s Figure 3, exhibiting a much steeper relationship between $\Sigma_{\rm SFR}$ and $\Sigma_{\rm gas}$. Runs I and E have substantially higher star formation rate surface densities at a given gas surface density than reported by \cite{2010ApJ...723.1019H}, but Run UF is rather closer. This is likely due to Run UF being an unbound cloud with, on average, lower star formation rates than the two bound clouds.\\
\indent In all three simulations (which are representative of the others), we see that the differences between the control simulation at either $t_{\rm ion}$ or $t_{\rm final}$ and the dual-feedback simulations at $t_{\rm final}$ are modest. All three sets of points occupy broadly the same region in the plots and there are no clear features to distinguish the control and dual--feedback calculations. It is thus difficult to discern the influence of feedback simply by analysing correlations between star formation rate surface density and gas surface density in this manner. This is in contrast to the results presented earlier where there is a clear effect of feedback on the correlation between stellar surface density and gas surface density. Star formation occurs in dense gas regardless of whether feedback is active or not but, as we saw in earlier sections, not all the dense gas in the feedback simulations is active in forming stars. Dense gas which is \emph{not} forming stars is obviously going to be poorly represented in plots constructed in the manner of Figure \ref{fig:sfr_sigma}, based on YSO counts.\\
\subsection{Unbinding of stars and clusters}
\begin{figure*}
     \centering
   \subfloat[Gas]{\includegraphics[width=0.33\textwidth]{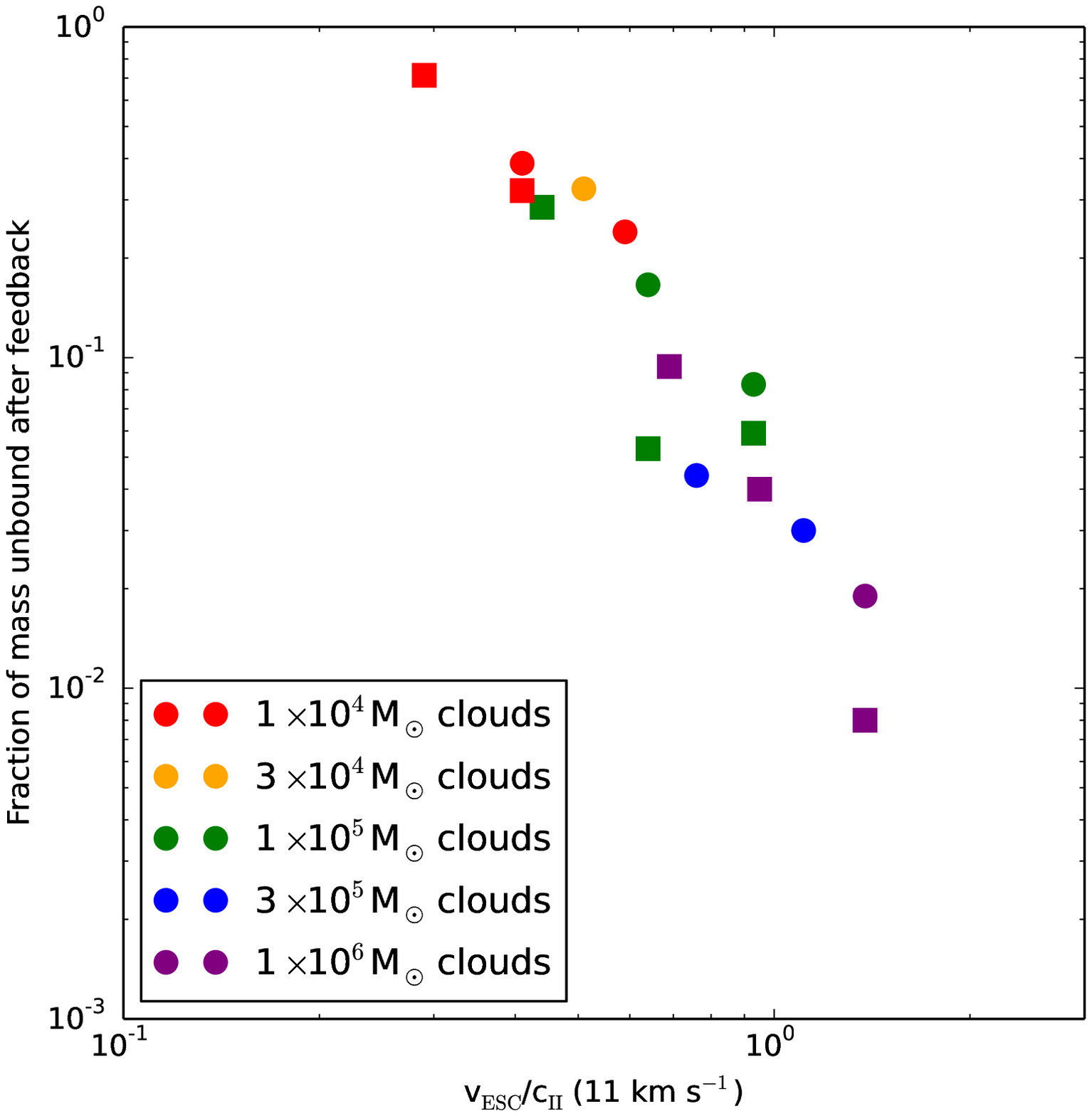}}
   \hspace{-0.05in}
      \subfloat[Stars (by number)]{\includegraphics[width=0.33\textwidth]{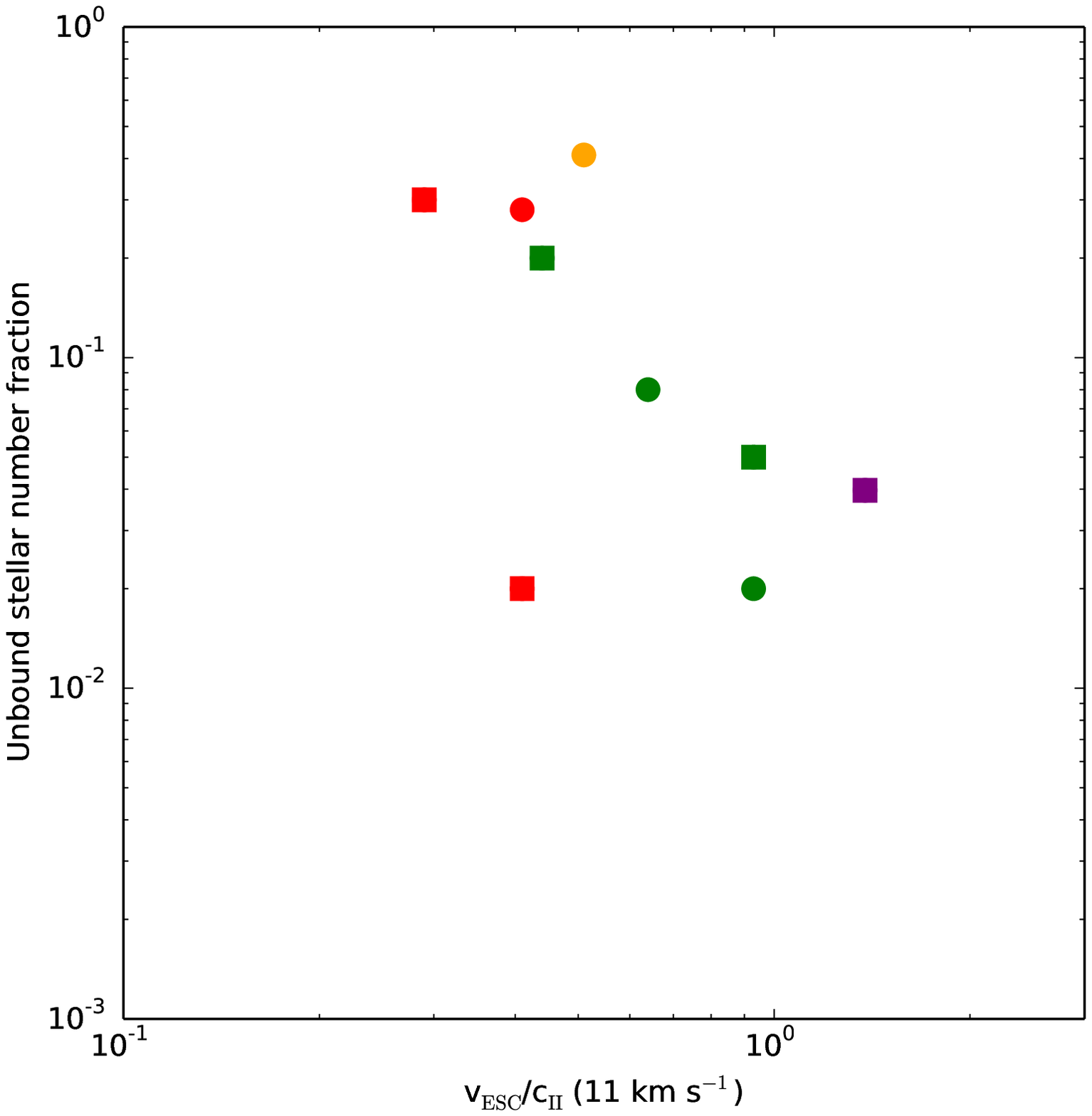}}
         \hspace{-0.05in}
      \subfloat[Stars (by mass)]{\includegraphics[width=0.33\textwidth]{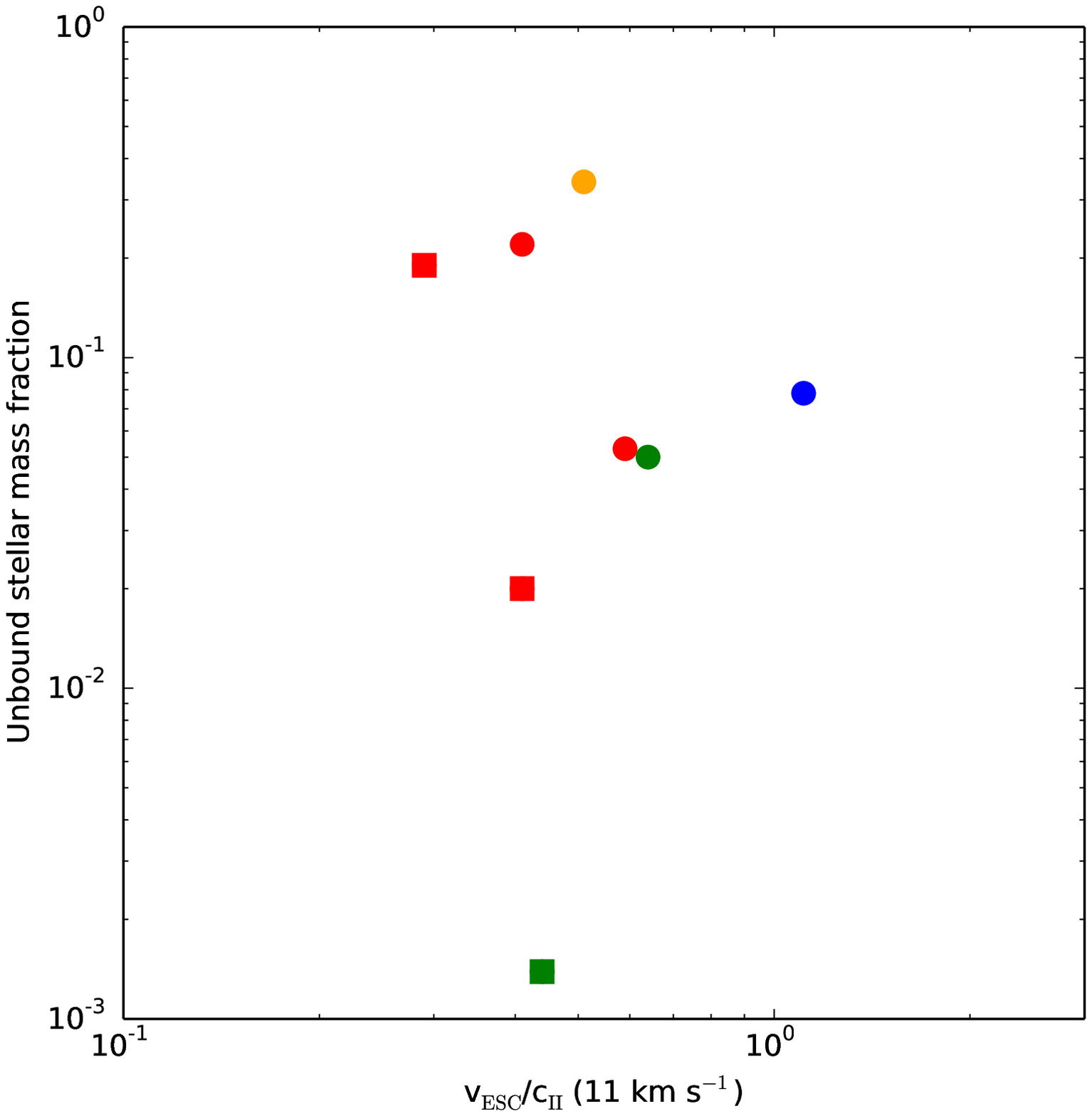}}
\caption{Unbound gas mass fractions (left panel), stellar number fractions (centre panel) and stellar mass fractions (right panel) plotted against cloud escape velocities normalised to the ionised sound speed (11 km s$^{-1}$) for all simulations. Colours denote cloud masses, circles are for initially unbound clouds and squares are initially bound clouds. In all cases, the relevant fractions from the control simulations have been subtracted, so not all simulations appear on all plots.}
\label{fig:unbound}
\end{figure*}

In Figure \ref{fig:unbound}, we show the gas mass fraction and mass and number fractions of stars unbound at the ends of our feedback calculations as a function of cloud escape velocities, expressed in units of the sound speed inside the HII regions (11 km s$^{-1}$). The relevant unbound fractions from the control simulations have been subtracted to isolate the effects of feedback, so the plots effectively show the quantities of material unbound \emph{by the action of feedback}. In the case of clouds where the mass or number fractions of stars unbound is very small, the mass or number fractions unbound in the control simulations may be slightly greater than in the feedback simulation. Therefore, not all simulations appear in the centre and right panels of the figure.\\
\begin{figure}
\includegraphics[width=0.48\textwidth]{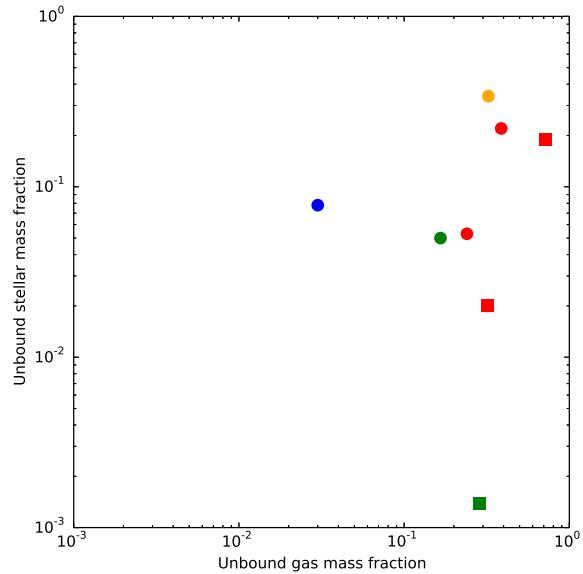}
\caption{Unbound stellar mass fractions plotted against unbound gas mass fractions. Colours again denote cloud masses, circles are for initially unbound clouds and squares are initially bound clouds. In all cases, the relevant fractions from the control simulations have been subtracted, so not all simulations appear in this plot.}
\label{fig:unbnd_gas_stars}
\end{figure}
\indent There is a tight correlation between the unbound gas mass and the cloud escape velocity, as discussed in \cite{2013MNRAS.430..234D}. The correlation between the unbound stellar \emph{numbers} and the cloud escape velocity is weaker with more scatter, and several clouds absent, from having greater numbers of stars unbound in the control simulations that in the dual--feedback run. There appears to be no correlation at all between the unbound stellar \emph{mass} and the cloud escape velocity. We check this in Figure \ref{fig:unbnd_gas_stars}, where we show that the unbound stellar mass fraction is apparently uncorrelated with but generally lower than, the unbound gas mass fraction, so that combined photoionisation and wind feedback is generally much less effective at unbinding stars than it is in unbinding gas. We explore the reasons for this lack of correlation in the Discussion section.\\
\indent Figure \ref{fig:unbound} suggests that neither the disruption of the clouds nor that of the clusters is mass--independent. In the case of the clouds, one important cause of this, as discussed in \cite{2013MNRAS.430..234D}, is likely to be the cloud escape velocity, or binding energy. In order to unbind the clouds, one must either supply sufficient momentum $p_{\rm crit}$ to accelerate the whole cloud beyond its escape velocity, or energy $E_{\rm crit}$ in excess of its binding energy. Since the clouds all have roughly the same surface densities, $M\sim R^{2}$, so $v_{\rm ESC}\sim M^{0.25}$, $p_{\rm crit}\sim M^{1.25}$, and $E_{\rm crit}\sim M^{1.5}$. However, as also mentioned in our earlier paper, the ability of the feedback--driven bubbles to explore the cloud volume also strongly affects their ability to destroy the clouds. At the ionised sound speed, an HII region can expand at most $\approx30$pc in the 3 Myr time window we have allowed here before the detonation of supernovae, which is greater than or comparable to the radii of the lower mass clouds, but much smaller than the radii of the more massive clouds.\\
\indent The vulnerability of clouds to feedback should also depend on their stellar content. As discussed in \cite{2014MNRAS.442..694D} and shown in the leftmost panel of Figure \ref{fig:mass_in_ostars}, with the exception of the nearly gas--exhausted Run F, the final star formation efficiencies of the simulations span about one decade from 1.3 to 11$\%$. Computing for each simulation the fraction of the cloud mass in stars above $20M_{\odot}$ (by simply counting in the 10$^{4}$ and 3$\times10^{4}$M$_{\odot}$ clouds and by assuming each subcluster in the more massive clouds has a Salpeter mass function between 0.5 and 100 M$_{\odot}$ and integrating for the more massive clouds) reveals a similar spread across the parameter space (centre--left panel of Figure \ref{fig:mass_in_ostars}).\\
\indent From the point of view of feedback, a more important quantity is the ionising luminosity per unit mass of the clouds (since ionisation dominates over winds), shown in the centre--right panel of Figure \ref{fig:mass_in_ostars}, which hints that the more massive clouds have smaller ionising luminosities per unit mass, although again with large dispersions, which is partly accounted for by the unbound clouds having generally lower star formation efficiencies. However, it  may also be due to the simplistic method used to assign feedback luminosities to the clouds of masses 10$^{5}$M$_{\odot}$ and above necessitated by our inability to resolve individual stars in these runs. This underestimates the luminosities of the most massive subclusters because it treats 30 M$_{\odot}$ stars as the basic unit of feedback, ignoring possible contributions from rare but much brighter more massive objects.\\
\indent However, we showed in \cite{2012MNRAS.424..377D} that the results of simulations are not sensitive to changes of  factors of several in the ionising luminosity. As we show in the rightmost panel of Figure \ref{fig:mass_in_ostars}, the dispersion in the fraction of clouds which are ionised is quite small with all calculations (save Run F) having between 1 and 12$\%$ of their mass ionised. Inspired by this, we presented in \cite{2013MNRAS.430..234D} a simple model treating the HII regions as pistons and considering what fraction of the cloud mass could be raised by their expansion to the appropriate escape velocity, finding a strong dependence of this fraction on the escape velocity itself.\\
\begin{figure*}
     \centering
   \subfloat[]{\includegraphics[width=0.255\textwidth]{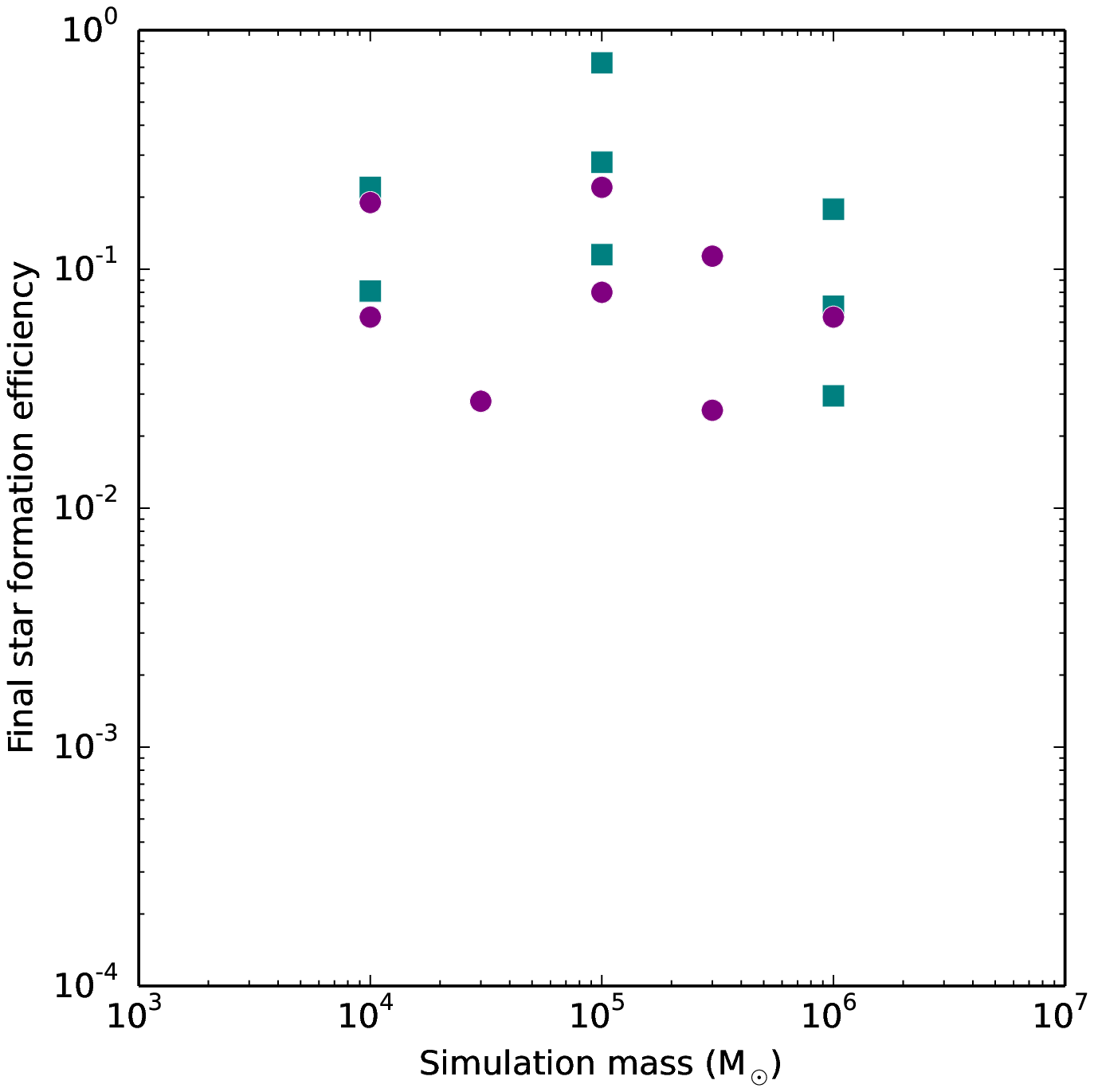}}
   \hspace{-0.12in}
      \subfloat[]{\includegraphics[width=0.255\textwidth]{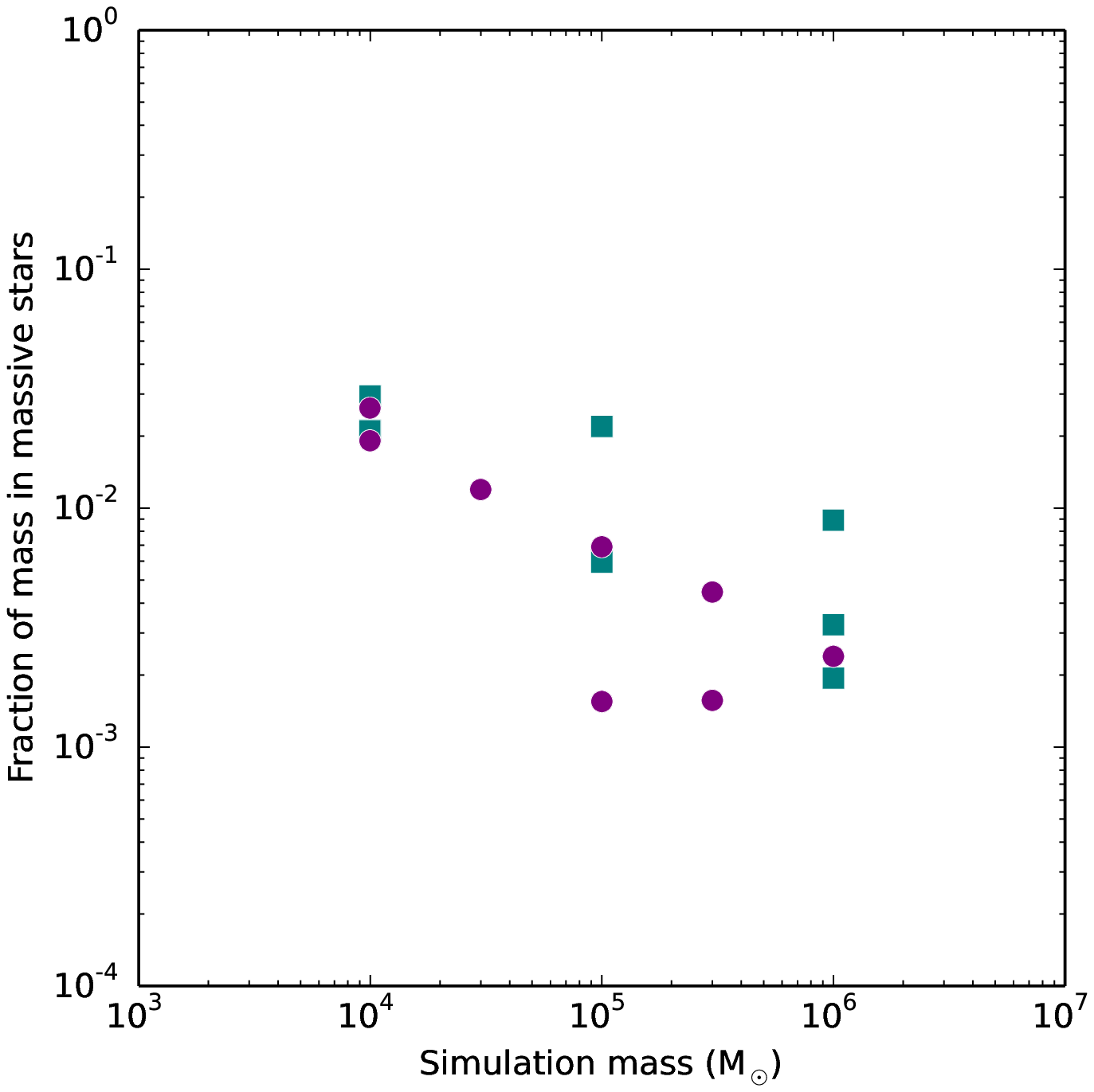}}
         \hspace{-0.12in}
      \subfloat[]{\includegraphics[width=0.255\textwidth]{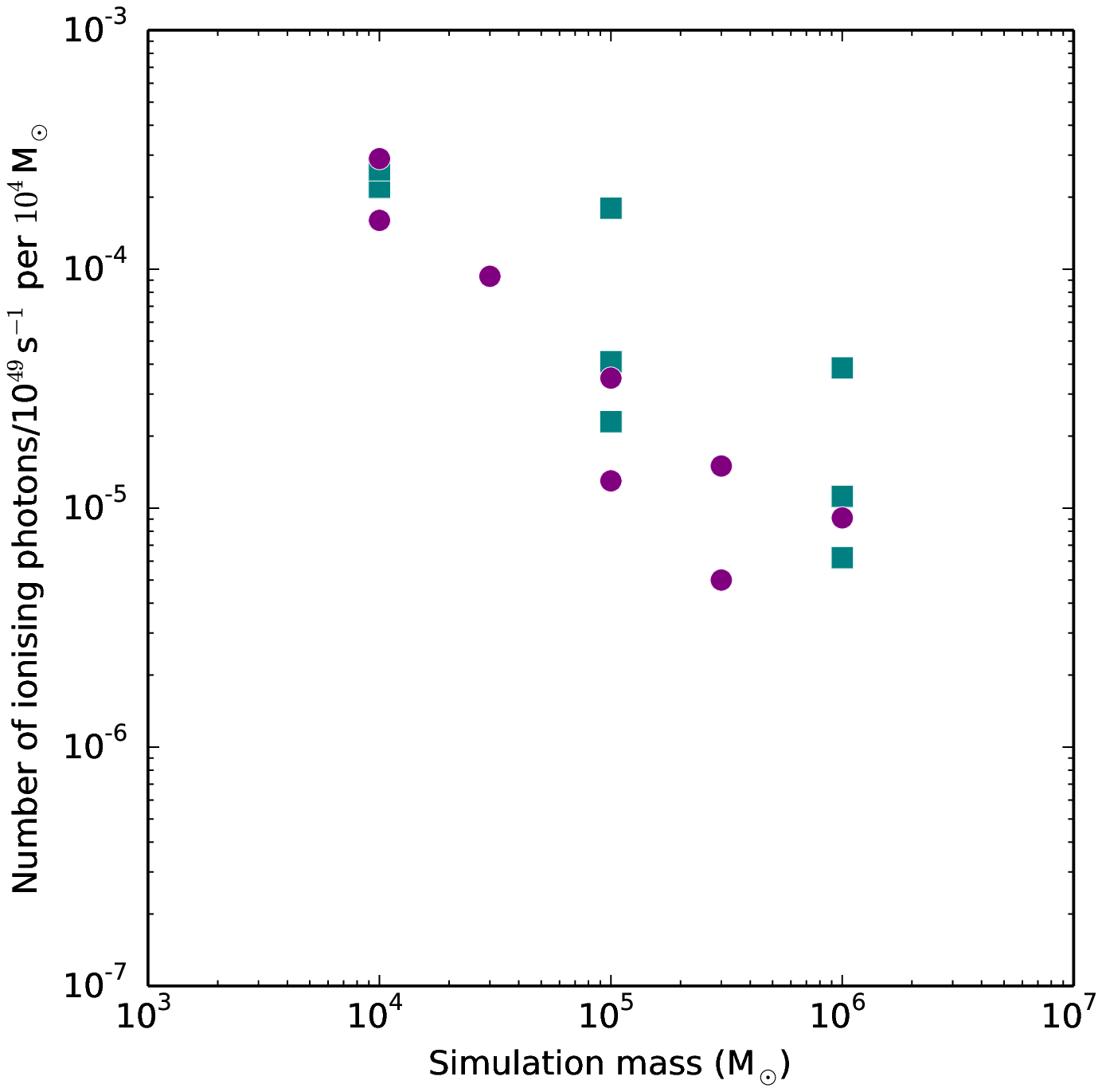}}
               \hspace{-0.12in}
      \subfloat[]{\includegraphics[width=0.255\textwidth]{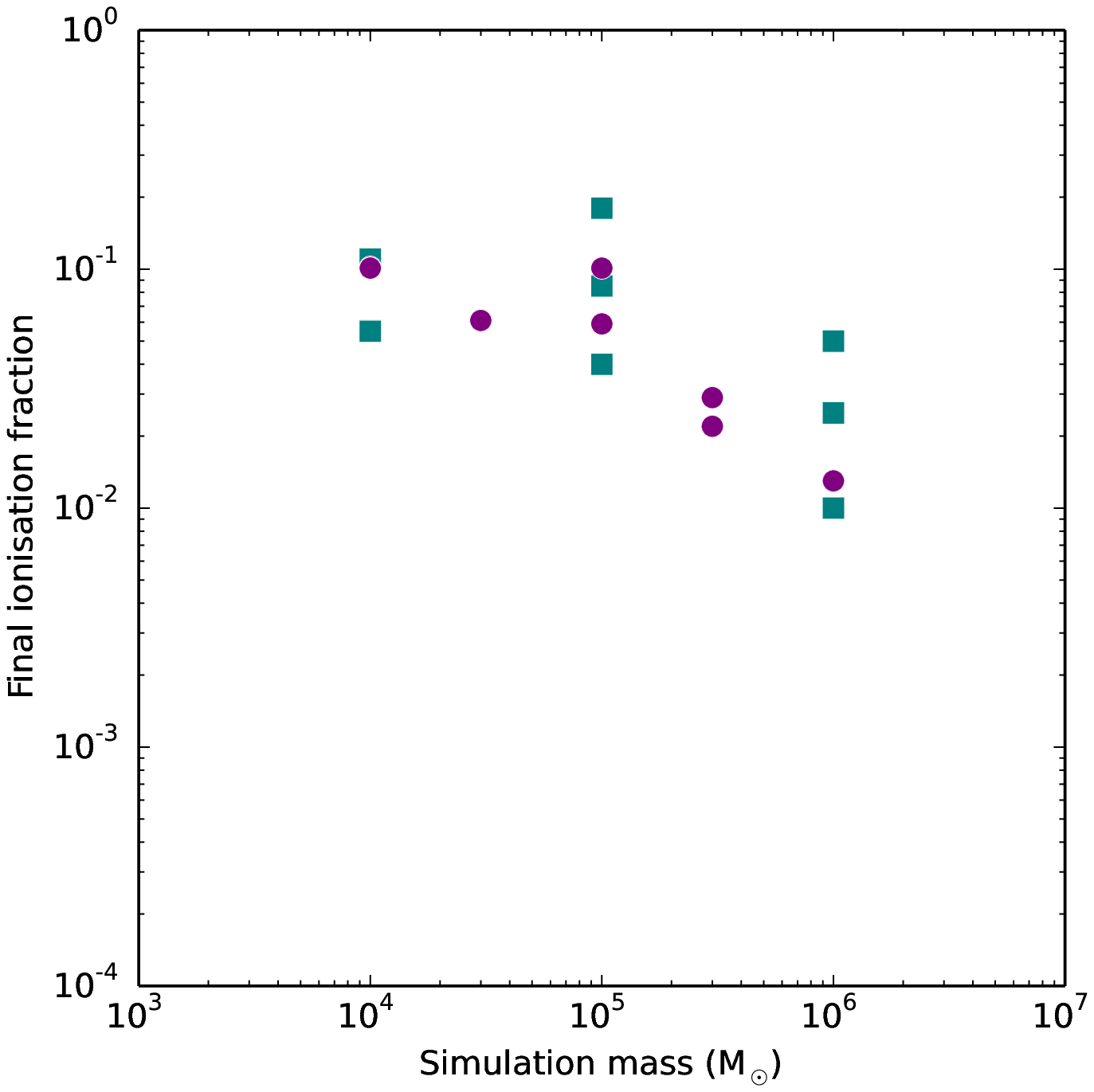}}
\caption{Final star formation efficiencies (leftmost panel), fraction of simulation mass in O--stars (centre--left panel), numbers of ionising photons per $10^{4}$M$_{\odot}$ (centre--right panel) and final global ionisation fraction (rightmost panel) for the final state of all simulations, with initially bound clouds shown as teal squares and initially unbound clouds as purple circles}
\label{fig:mass_in_ostars}
\end{figure*}
\subsection{Cluster emergence}
\begin{figure}
\includegraphics[width=0.48\textwidth]{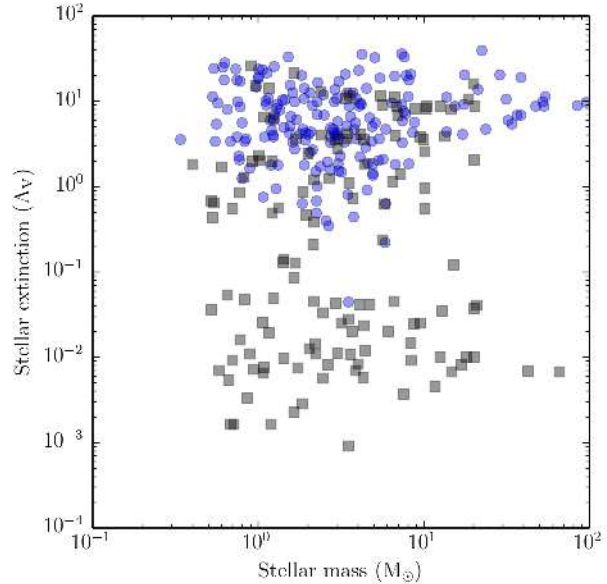}
\caption{Foreground extinctions in V--magnitudes for all stars in the control (blue) and dual--feedback (grey) as a function of stellar mass in Run I at the ends of the simulations, at times of 7.56Myr when viewing the system along the $z$--axis.}
\label{fig:extinct}
\end{figure}

\begin{figure*}
     \centering
   \subfloat[Run I]{\includegraphics[width=0.33\textwidth]{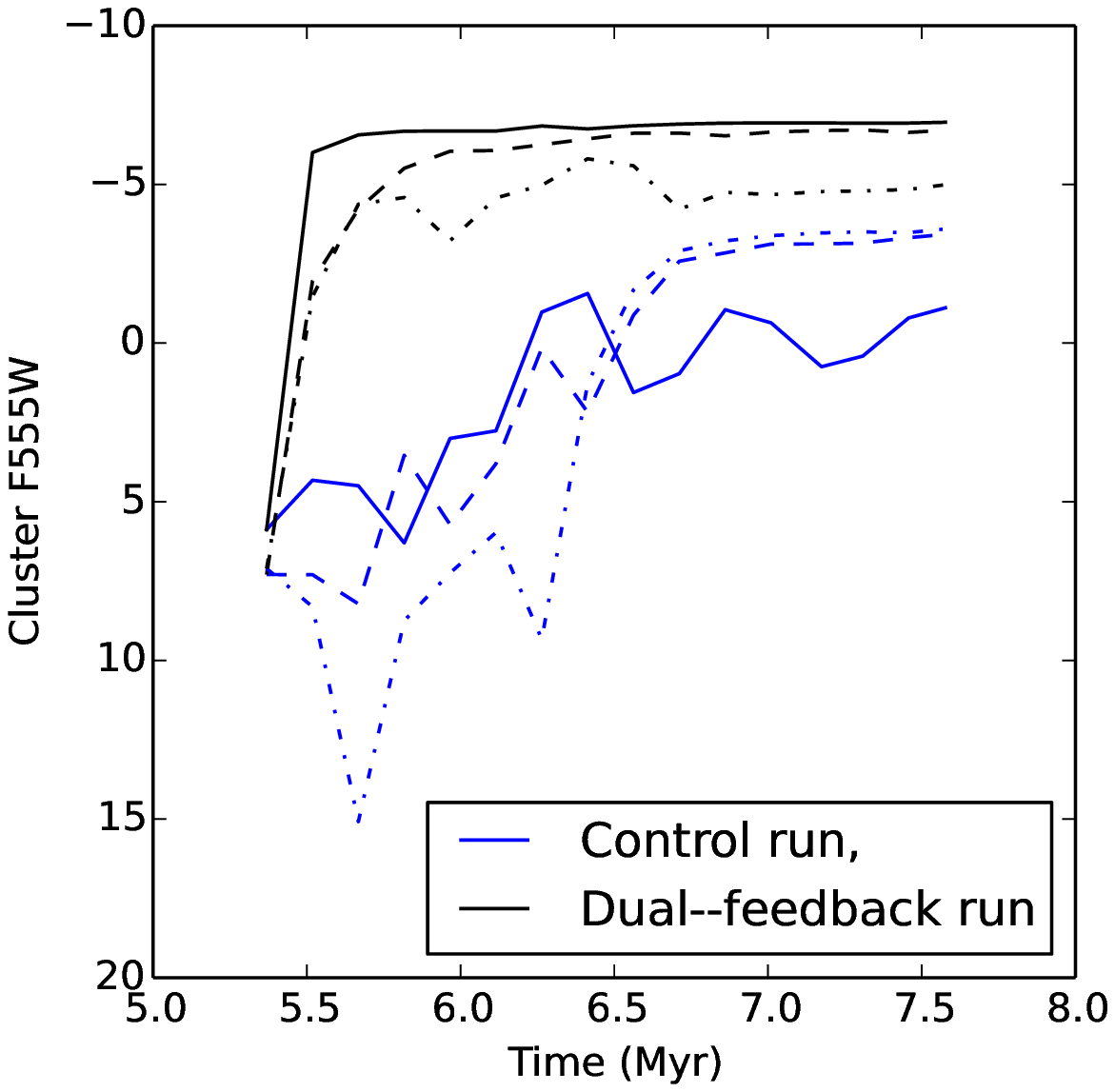}}
   \hspace{-0.05in}
      \subfloat[Run J]{\includegraphics[width=0.33\textwidth]{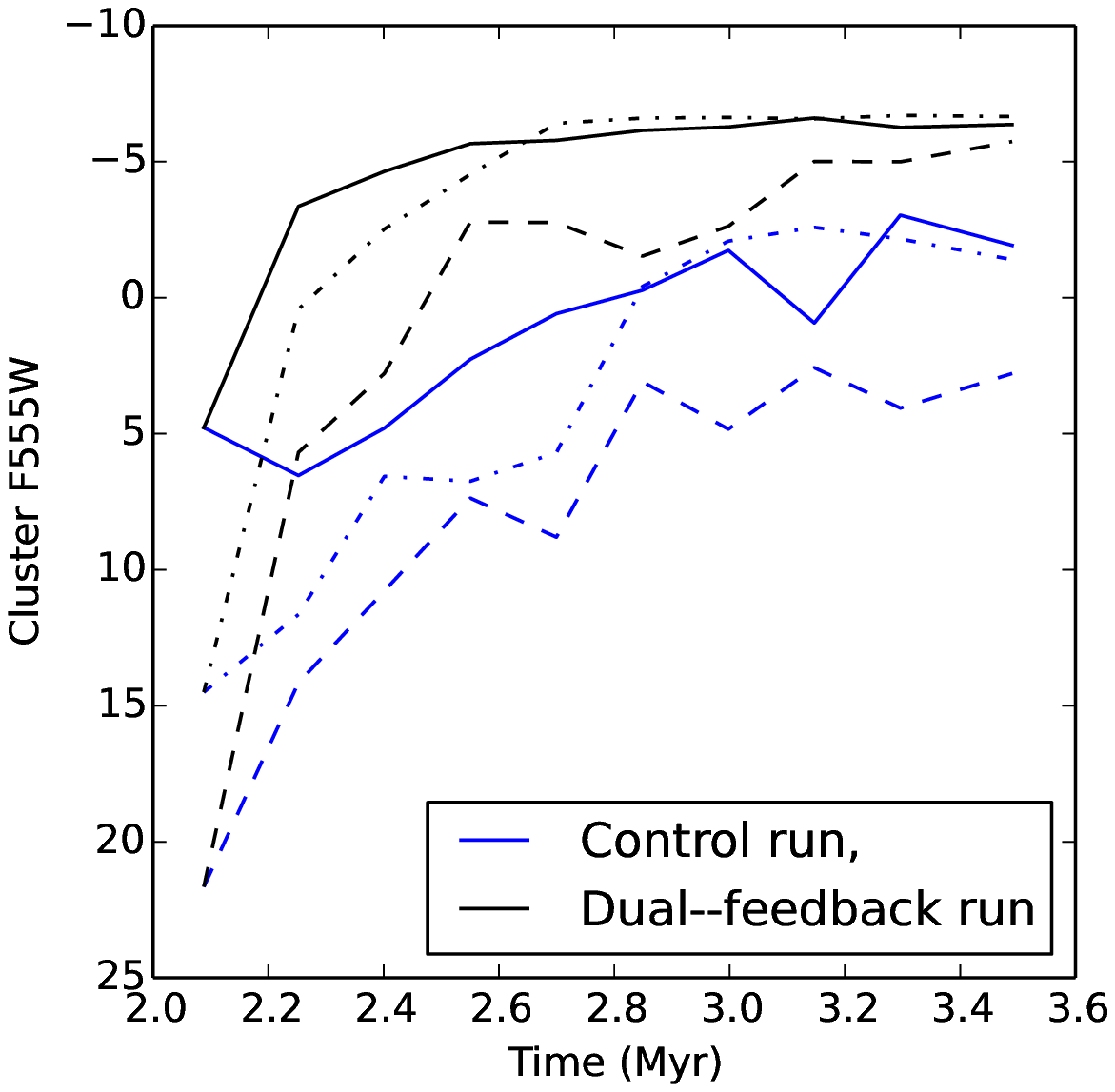}}
         \hspace{-0.05in}
      \subfloat[Run UQ]{\includegraphics[width=0.33\textwidth]{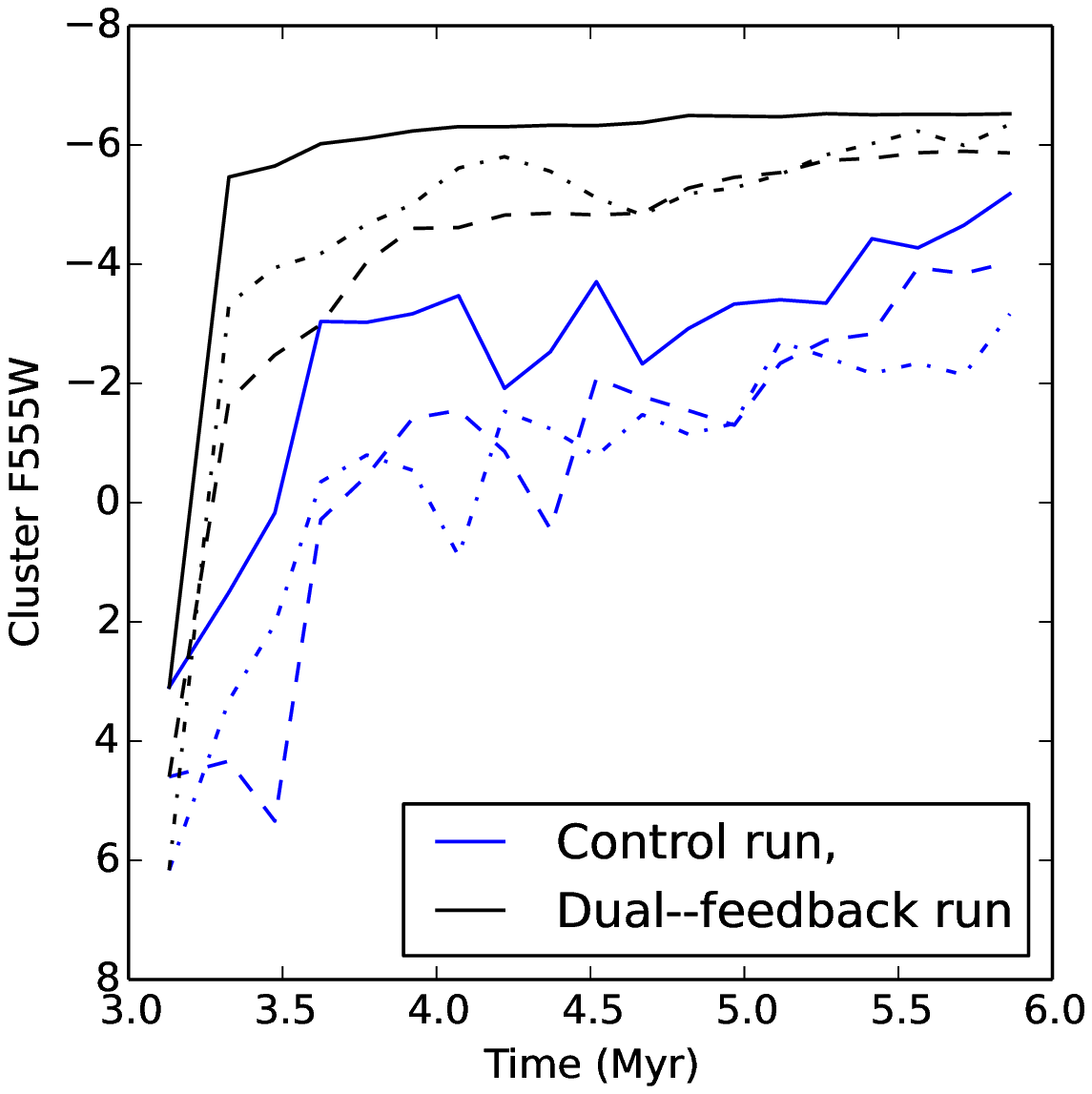}}
\caption{Absolute V--band magnitudes of clusters as a function of time in Runs I (left panel), J (centre panel) and UQ (right panel). Control simulations are shown as blue lines and dual--feedback runs as black lines. Each cluster is observed along the $x$-- (dash--dot lines), $y$-- (dashed lines) and $z$--axis (solid lines).}
\label{fig:reveal}
\end{figure*}

\begin{figure}
\includegraphics[width=0.48\textwidth]{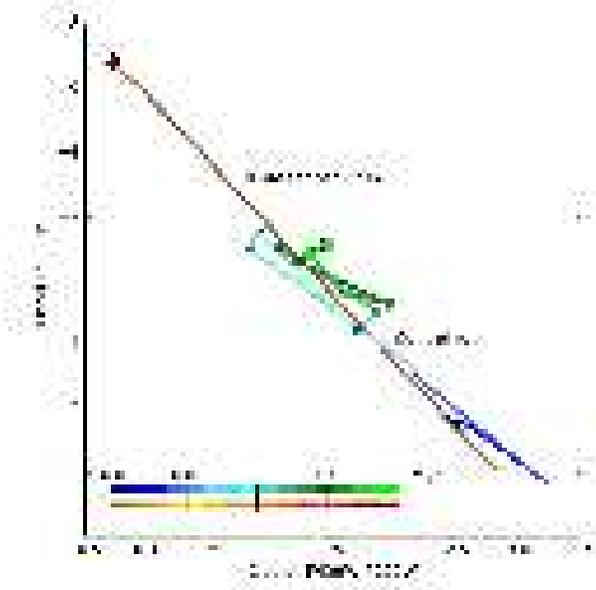}
\caption{Evolution of the colour--magnitude diagram of the clusters from the control (blue--green colour map and circles) and dual--feedback (yellow--red colour map and triangles) Run I calculations in the F555W versus F439W-F555W space. The elapsed time since the point when feedback is enabled in the dual--feedback run is given in the colour bars at bottom left.}
\label{fig:runi_colours}
\end{figure}

\begin{figure}
\includegraphics[width=0.48\textwidth]{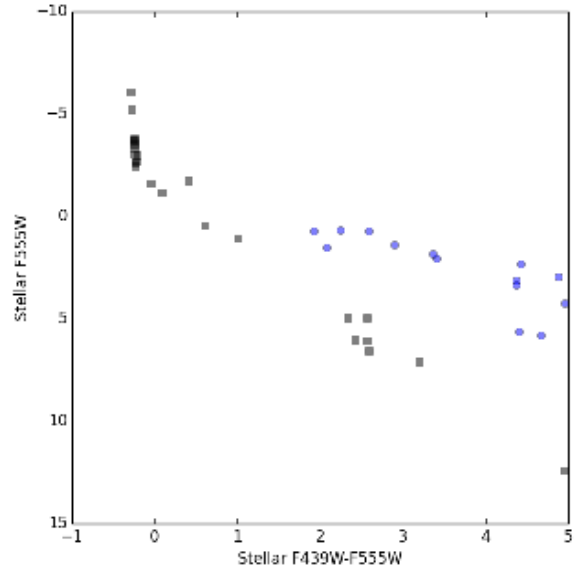}
\caption{Colour--magnitude diagram of individual stars in the control (blue circles) and dual--feedback (black squares) Run I calculations in the F555W versus F439W-F555W space at the endpoints of the two calculations.}
\label{fig:runi_colours_s}
\end{figure}

The observed emergence of stellar clusters from their embedded phase occurs as a result of gas being cleared from the  clusters, regardless of whether the loss of gas dynamically impacts the clusters. A detailed numerical study of this phenomenon strictly requires advanced radiation transport calculations which are beyond the scope of this work. However, some insight can be gained by a simple analysis.\\
\indent Open clusters and stellar associations are generally easily visible in the optical (e.g. the V--band), whereas embedded clusters typically lie behind 5--100 magnitudes of extinction in the V--band \citep{2003ARA&A..41...57L} and are rendered nearly invisible at optical wavelengths. It is the transition between these  states which is generally referred to as `emergence'. Infrared (e.g. K--band) extinctions are around eight times smaller than optical extinctions for a given column density, so embedded clusters obscured in the optical may be easily visible in the infrared.\\
\indent Protostars and pre--main--sequence objects are very bright in the infrared thanks to their accretion luminosities and the reprocessing of radiation by their dust--laden gaseous envelopes. Modelling the K--band luminosities of such objects is very difficult and we will not attempt it here. The visible and ultraviolet light, however, should be dominated by the bright massive stars, which become well settled on the main sequence while still embedded. We therefore choose to treat the massive stars in our clouds as having main sequence colours and luminosities and examine the extinction of their UV/visible light by the remaining gas in the simulations. We then sum the contribution of the massive stars to estimate the total visible magnitude of the whole stellar population, as it would appear if it were an unresolved cluster.\\
\indent We utilise tabulated colours computed for stars in Hubble Space Telescope filters from \cite{2001A&A...366..538L}. We choose solar metallicity and 1Myr isochrones as a reasonable average age for our stars. We concentrate on the F336W (approximating the U--band), F439W (the B--band) and F555W (the V--band) filters. We include contributions from all stars more massive than 10M$_{\odot}$. We compute the intrinsic magnitude of each star from its mass only and obtain the extinction in front of each star from the column density between the star and an observer at infinity in the $x$--, $y$-- or $z$--directions. The column densities are converted to extinctions by assuming that U--, B-- and V--band extinction laws apply, so that
$A_{\rm V}$=$N_{\rm H}$/10$^{21}$cm$^{-3}$, $A_{\rm B}$=1.322$A_{\rm V}$ (this corresponds to the canonical mean extinction law \citep[e.g.][]{1999PASP..111...63F}, but we note that values as low as $A_{\rm B}$=1.322$A_{\rm V}$ may be appropriate for dense regions of the ISM) and $A_{\rm U}$=1.581$A_{\rm V}$ \citep{1979ARA&A..17...73S}.\\
\indent Figure \ref{fig:extinct} shows, for the final timesteps of the control (blue) and dual--feedback (grey) Run I simulations, the values of $A_{\rm V}$ for \emph{all} stars when viewing the system along the $z$--axis, plotted as a function of stellar mass. The extinctions in the control run are sharply peaked at $A_{\rm V}\approx10$mag and extend up to $\approx$50 mag. Those in the dual--feedback run fall into two clear groups. One, comprising roughly half the stars and corresponding to the population of stars embedded in or near the bubble walls, has a mean extinction also close to 10 magnitudes. The other half have extinctions less then 1 mag and often negligible, corresponding to the exposed cluster inside the cleared out feedback--driven bubble. Apart from the few very most massive stars in the feedback simulation, whose extinctions are very low, there is little dependence of extinction on stellar mass in either simulation. In particular, in the feedback simulation, a star of any mass can have its environs completely evacuated of gas by the few massive stars present in the system.\\
\indent To compute the magnitudes of the \emph{clusters}, we first convert the magnitudes of each star to fluxes in arbitrary units using
\begin{equation}
F_{\rm clus,U,B,V}=\sum_{\rm stars}{F_{\rm *,U,B,V}}=\sum_{\rm stars}{F_{0}\times10^{-(M_{\rm U,B,V}-A_{\rm U,B,V})/2.5}},
\end{equation}
then sum over all sources and convert back to magnitudes using
\begin{eqnarray}
M_{\rm clus,U,B,V}=-2.5{\rm log}\left(\frac{F_{\rm clus,U,B,V}}{F_{0}}\right).
\end{eqnarray}
\indent In Figure \ref{fig:reveal}, we plot the total magnitude in the F555W filter as a function of time for the control (blue lines) and dual--feedback (black lines) Runs I (left), J (centre) and UQ (right) simulations. In each case, we plot results looking down the $x$-- (dash--dot lines), $y$-- (dashed lines) and $z$--axis (solid lines) to gain an idea of how projection effects may influence the results.\\
\indent The initial magnitudes for the three clusters are in the range 5--25. There is substantial variation in the results depending on the viewing angle. In Run J in particular, changing the viewing angle results in variations of almost 20 magnitudes in the cluster brightness. This is purely due to the highly anisotropic nature of the dense gas in the which the majority of the stars are initially embedded. However, the same qualitative conclusions may be drawn from all three viewing angles.\\
\indent In both the control and dual--feedback simulations, the clusters generally become brighter with time, often quite rapidly. In the control simulations, this is due largely to unabated accretion producing brighter and brighter stars. The increase in brightness is not smooth, however, with considerable jumps visible. These features are a result of the non--steady delivery of gas to the core regions of the cloud by accretion flows along the filaments, leading to large excursions in extinction. In the dual--feedback simulations, by contrast, the increase in cluster brightness is generally much more abrupt, with all three of these simulations producing exposed clusters with brightnesses of -6-- -7 mag in the F555W filter, with differences in viewing angle resulting in spreads of about 1 mag. However, the timescales on which the clusters are revealed also depend somewhat on viewing angle. In Run J, the cluster as viewed along the $z$--axis appears to achieve a magnitude brighter than zero only $\approx$10$^{5}$yr after ionisation is enabled, whereas this timescale is $\approx$4$\times$10$^{5}$yr for an observer on the $y$--axis. The greater variation of this timescale in Run J is due to the less effective clearing out of the cluster volume by feedback in this calculation relative to the other two shown.\\
\indent Figure \ref{fig:runi_colours} shows the evolution of the cluster colour--magnitude diagram in the F555W versus F439W-F555W space of the control (blue--green colour map and circles) and dual--feedback (yellow--red colour map and triangles) Run I, as viewed along the $z$--axis, with the colour bar at bottom left giving the elapsed time since feedback was enabled in the dual--feedback calculation. The control simulation becomes brighter and bluer but, at later times, oscillates about a colour of $\approx1.5$ and a magnitude of $\approx0$. By contrast, the dual--feedback simulation moves very rapidly from the bottom right to the top left of the CMD, quickly acquiring a stable colour of $\approx-0.3$ and stable magnitude of $\approx-7$.\\
\indent Figure \ref{fig:runi_colours_s} shows a CMD for the individual stars ($>10$M$_{\odot}$) in the control and dual--feedback Run I simulations at the ends of the runs. All of the stars in the control simulation are strongly reddened and control run stars of a given colour are several magnitudes brighter than stars of the same colour from the dual--feedback simulation. The stars from the control simulation are intrinsically brighter, since they are able to acquire larger masses, but are buried in much larger extinctions, so appear redder. The population embedded in the bubble walls in the dual--feedback simulation is visible at a colour of 2--3 and a magnitude of $\approx5$. The exposed population is visible at the top left as a vertical grouping characteristic of massive stars on the main sequence, whose colours vary little but whose magnitudes vary substantially.\\
\indent The timescale on which clusters become observable is of crucial observational importance and can be estimated using the above analysis. In Table \ref{tab:tminus5}, we give, for Runs I, J, UF, UP and UQ the time at which star formation began ($t_{\rm SF}$), the time at which feedback was enabled ($\tau_{\rm FB}$), the time at which the clusters became brighter than V=-5 ($t_{\rm -5}$), chosen in part because no cluster in any control simulation achieves this brightness) and the time of the end of the simulations ($t_{\rm end}$). We use these figures to compute, at the endpoints of the simulations, the total duration of star formation $\tau_{\rm SF}$, the time since feedback was enabled $\tau_{\rm FB}$, the time for which the cluster in the feedback run has been brighter than a V--magnitude of -5 $\tau_{\rm -5}$ , the fraction of the star--forming lifetime for which the cluster is brighter than V=-5, f$_{\rm SF}$, and the fraction of the time for which the O--stars have been active for which the cluster is brighter than V=-5, f$_{\rm FB}$. It is clear from Figure \ref{fig:runi_colours} that $t_{\rm -5}$ varies substantially with viewing angle. We show this crudely by giving for each cloud the median value of $t_{\rm -5}$ from the $x$--, $y$-- and $z$--projections, with the other two as rough upper and lower limits.\\
\indent As alluded to previously, the variation with viewing angle in the times at which the clusters become brighter then V=-5 can be substantial, with variations up to almost 1Myr in the case of Run UQ. However, the time delay between the onset of star formation and the formation of the first few O--stars, defining when feedback is enabled, is also approximately 1Myr in all simulations, and the star formation timescales are in the range 2--4 Myr in total (and would be longer in the case of Runs J and UP if it had been practical to continue the simulations for the full 3Myr after the initiation of feedback). The fractions of the star formation timescales for which the clusters are brighter than V=-5 are in the range 39--54$\%$ with typical variations of $\approx15\%$ and absolute maxima and minima 69$\%$ and 23$\%$ respectively. Clusters should thus be bright for about half the time interval between the onset of star formation and the detonation of the first supernovae, but the scatter in this quantity is significant. Naturally, the fraction of time since the O--stars formed for which the clusters are this bright is higher, between 59 and 84$\%$ with absolute maxima and minima of 100$\%$ and $25\%$. The variations in this quantity are thus even larger. These variations imply that substantial fractions of cluster in magnitude--limited surveys may be missed purely through inhomogeneities in the intra--cluster gas.\\
\begin{table*} 
\begin{tabular}{|l|l|l|l|l|l|l|l|l|l|}
Run & $t_{\rm SF}$ & $t_{\rm FB}$ & $t_{\rm -5}$ & $t_{\rm end}$ & $\tau_{\rm SF}$ & $\tau_{\rm FB}$ & $\tau_{\rm -5}$ & f$_{\rm SF}$ & f$_{\rm FB}$\\
\hline
I &4.18&5.37&5.73$^{+0.68}_{-0.35}$&7.58&3.40&2.21&1.85$^{+0.35}_{-0.68}$&0.54$^{+0.11}_{-0.20}$&0.84$^{+0.15}_{-0.31}$\\
\hline
J &1.34&2.09&2.66$^{+0.34}_{-0.16}$&3.49&2.15&1.40&0.83$^{+0.16}_{-0.34}$&0.39$^{+0.07}_{-0.16}$&0.59$^{+0.12}_{-0.24}$\\
\hline
UF &2.08&3.28&4.43$^{+0.13}_{-0.81}$&6.23&4.15&2.95&1.80$^{+0.81}_{-0.13}$&0.43$^{+0.20}_{-0.03}$&0.61$^{+0.17}_{-0.04}$\\
\hline
UP &0.92&1.83&2.60$^{+0.46}_{-0.72}$&3.71&2.69&1.88&1.11$^{+0.72}_{-0.46}$&0.41$^{+0.17}_{-0.15}$&0.59$^{+0.38}_{-0.24}$\\
\hline
UQ &1.94&3.13&3.78$^{+0.99}_{-0.64}$&5.86&3.92&2.73&2.08$^{+0.64}_{-0.99}$&0.53$^{+0.16}_{-0.26}$&0.76$^{+0.24}_{-0.37}$\\
\end{tabular}
\caption{Comparison of the times (all in Myr) at which star formation begins ($t_{\rm SF}$), the time at which feedback is enabled ($t_{\rm FB}$), the time at which the clusters reach V=-5 ($t_{\rm -5}$), the end time of each simulation ($t_{\rm end}$), the total duration of star formation $\tau_{\rm SF}$, the time since feedback was enabled $\tau_{\rm FB}$, the time for which the cluster in the feedback run has been brighter than a V--magnitude of -5 $\tau_{\rm -5}$, the fraction of time since star formation began for which the clusters have been brighter than V=-5 (f$_{\rm SF}$) and the fraction of the time since feedback was enabled for which the clusters have been brighter than V=-5 (f$_{\rm FB}$), for Runs I, J, UF, UP, UQ.}
\label{tab:tminus5}
\end{table*}
\section{Discussion}
\subsection{Disruption of clouds and clusters}
\indent We found that the disruption of the clouds by feedback is not mass--independent and that the disruption of the stellar clusters is also not mass--independent, and is substantially less severe. This result differs from previous work on this issue \citep[e.g][]{1978A&A....70...57T,2003MNRAS.338..673B,2006MNRAS.373..752G,2007MNRAS.380.1589B,2013A&A...555A.135P}. These authors model the effects of gas expulsion by implicitly or explicitly allowing an N--body system to come into equilibrium with an artificial potential, then removing that potential instantaneously or on a prescribed timescale. The stellar system modelled is initially smooth, often in the form of a Plummer sphere, and the background potential is also smooth. The contribution of the stellar mass to the total mass density (the `star formation efficiency') is taken to be the same at all radii.\\
\begin{figure}
\includegraphics[width=0.48\textwidth]{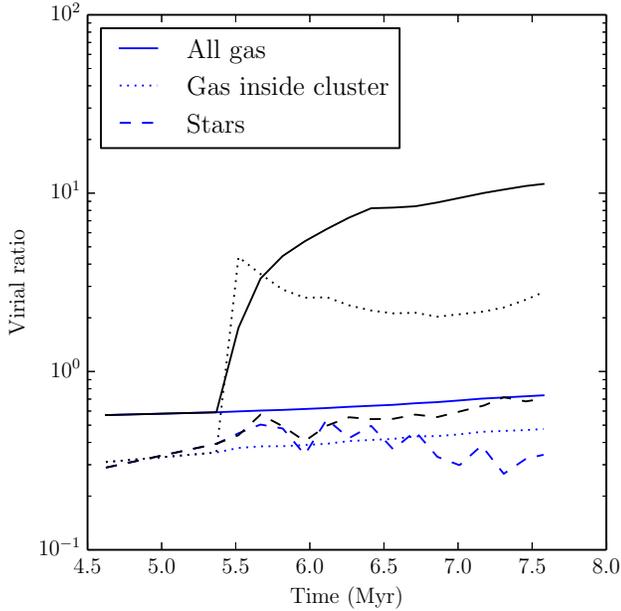}
\caption{Virial ratios of all cold gas (solid lines), the gas occupying the volume defined by the location of the most massive star and the furthest star from this point (doted lines), and of all stars (dashed lines) in the control (blue lines) and dual--feedback (black lines) Run I calculations.}
\label{fig:vratios}
\end{figure}
\begin{figure}
\includegraphics[width=0.48\textwidth]{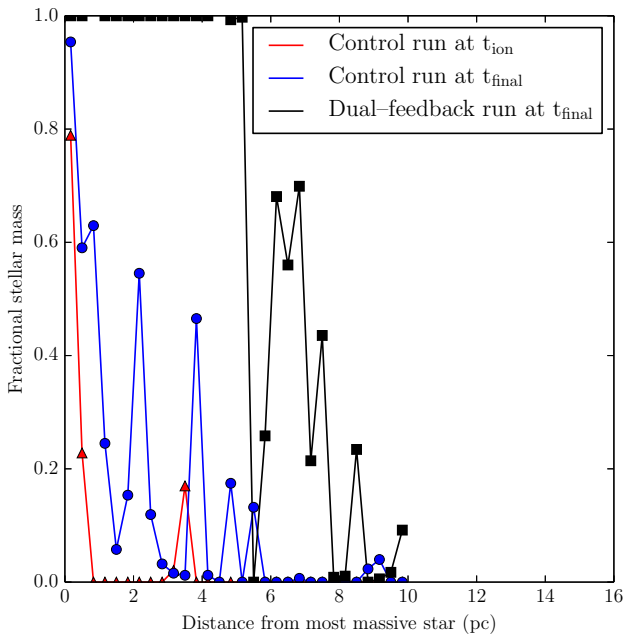}
\caption{Fractional contribution of stellar mass to the total mass in radial bins as a function of radius from the most massive star at the onset of ionisation (red line and triangles), the end of the control run (blue line and circles) and at the end of the dual--feedback (black line and squares) in Run I.}
\label{fig:sfer}
\end{figure}
\indent These are tempting simplifications, but hydrodynamic simulations in which star formation is self--consistently modelled have shown that they are not correct. \cite{2009ApJ...704L.124O,2012MNRAS.419..841K,2012MNRAS.420..613G} all find that the stellar and gaseous components of embedded clusters are \emph{never} in virial equilibrium with each other, and that the stars tend to be subvirial in both the absolute sense of having a virial ratio less than 0.5, and a relative sense in having a smaller virial ratio than that of the gas. The stars and gas thus form two largely decoupled systems.\\
\indent We find that the same is true in our simulations. Figure \ref{fig:vratios} depicts, for Run I, the virial ratio of all cold gas (solid lines), the cold gas occupying the same volume as the stars (defined by the smallest sphere centred on the most massive which contains all the stars, dotted lines), and of all stars (dashed lines) in the control (blue lines) and dual--feedback (black lines) calculations.\\
\indent We here define the virial ratio as a simple ratio of kinetic energy to gravitational potential energy $E_{\rm kin}/|E_{\rm grav}|$. The kinetic energy for each component (e.g. cold gas, stars) is computed by summing the kinetic energies of each individual star or gas particle in the centre of mass frame of the whole simulation. The potential energy is computed by summing over all star or gas particles the total potential energy generated by all mass components of the simulation on each particle.\\
\indent The stellar virial ratio is always less than that of the gas in both calculations. In the control simulation, the virial ratios of both components change only modestly over the course of the calculation, the stars becoming more subvirial, while the gas slowly approaches a virial ratio of unity, likely due to the slow expansion of the outer regions of the cloud. In the feedback simulation, the virial ratio of the cold gas increases strongly and rapidly, but that of the stars increases much more slowly.\\
\indent If only gas inside the volume occupied by the stars is considered, the picture is slightly more complex. Initially, the virial ratios of the stars and cospatial gas are very similar. When feedback is enabled, the virial ratio of this gas in the dual--feedback simulation rises very sharply, since it is quickly and directly affected by feedback. As time progresses, the virial ratio of this material initially drops, before rising again at late times as the second--generation feedback sources become active. However, at all times after feedback is enabled, this gas has a substantially higher virial ratio than the stars. In the control simulation, the virial ratio of the gas cospatial with the stars simply rises gradually with time. Since the stars become more subvirial, the intracluster cluster becomes somewhat supervirial with respect to the stars.\\
\indent The virial states of the gas and stars in these simulations are never strongly coupled, even when only considering the gas cospatial with the stars. This is especially true of the dual--feedback simulation, where gas inside the cluster is always supervirial with respect to the stars by a factor of at least four. It is not surprising that the expulsion of even a large fraction of the surviving gas in some of the feedback simulations has only a modest effect on the dynamics of the stars.\\
\indent We reinforce this point in Figure \ref{fig:sfer}, which shows the fractional contribution of stellar mass to the total mass in radial bins as a function of radius from the most massive star in the control run at the onset of ionisation (red), at the end of the control run (blue) and at the end of the dual--feedback Run I (black). The total star formation efficiencies averaged over the entire clouds are, respectively, 4, 13 and 8 percent. However, the local star formation efficiencies clearly depart strongly from the average. This is particularly true in the dual--feedback simulation, where the clearing of gas from the central cluster and its expansion result in a region $\approx$6pc in radius where all the mass is stellar. However, even in the absence of feedback, the local star formation efficiencies in the much more compact central cluster both at the onset of ionisation and at the end of the control simulation are much higher than the global average. The assumption that the gas and stars are evenly mixed is therefore a poor one. The local star formation efficiency in the central cluster at the onset of ionisation is well in excess of the critical value of 33 percent identified by \cite{2007MNRAS.380.1589B} as being the minimum required to survive \emph{adiabatic} gas expulsion, so the survival of this stellar system in the dual--feedback simulation is not, in fact, surprising. Although more than half of the cloud mass is unbound by feedback, the gas is mostly lost from regions where there are no stars.\\
\indent Recently, more advanced N--body simulations have improved on the methodology of early calculations. These enhancements have taken two main forms. \cite{2010MNRAS.404..721M,2012MNRAS.425..450M} dropped the assumption of smooth stellar distributions by extracting the sink particles formed in the simulations of (respectively) \cite{2009MNRAS.392..590B} and \cite{2011MNRAS.410.2339B}. By dispensing with the gas, they effectively assumed instantaneous gas expulsion, which is the most destructive to the remaining stellar system. \cite{2010MNRAS.404..721M} found that 30--40 percent of the stars from the \cite{2009MNRAS.392..590B} calculation remained bound after 10 Myr of post--gas expulsion evolution, despite the hydrodynamical calculation having a star formation efficiency of 38 percent, well below the canonical 50 percent required to survive \emph{instantaneous} gas loss \citep{1980ApJ...235..986H}. The results of \cite{2012MNRAS.425..450M} were qualitatively similar and for similar reasons -- the clusters formed in the calculation of \cite{2011MNRAS.410.2339B} were already gas--depleted and virialised, as reported by \cite{2012MNRAS.419..841K}.\\
\indent \cite{2013MNRAS.432..986P} and \cite{2015MNRAS.446.4278P} used a similar approach to continue a subset of the simulations of \cite{2012MNRAS.424..377D}, \cite{2013MNRAS.430..234D} and \cite{2014MNRAS.442..694D}. They assumed that, after the $\approx$3Myr action of ionisation feedback from the clusters' O--stars, the first supernova in each model cloud instantaneously expelled the remaining gas. By comparison with control simulations in which no pre--supernova feedback operated, they examined what effect the slow gas expulsion caused by ionising feedback might have on the evolution of the clusters after sudden but delayed gas expulsion. They found that it was difficult to identify clear trends in the effects of pre--supernova feedback, but that in all cases, the instantaneous removal of the remaining (usually large) fraction of gas left substantial fractions of the stars bound (more than half in all but one of the ten simulations analysed). The fractions of unbound stars in all calculations increased with time over the 10 Myr of N--body evolution, but this was a result of stellar--dynamical and stellar--evolutionary effects.\\ 
\indent Other authors have sought to improve the N--body modelling of young clusters by using more realistic, but still artificial, initial conditions. \cite{2011MNRAS.414.3036S} and \cite{2013MNRAS.428.1303S} construct artificial substructured clusters using either a fractal space--filling method, or by placing a number of small Plummer spheres within a larger Plummer potential. This has the effect of ensuring that the local star formation efficiency can be much higher than the mean efficiency, as we observed in Figure \ref{fig:sfer}, and these authors find that the local SFE is the main determinant of cluster survival, provided that gas expulsion occurs late enough that the stars have relaxed. However, the background potential representing the gas is still smooth and can still only be homologously removed.\\  
\subsection{Emergence of embedded clusters}
\indent For the low--mass clouds, the stellar populations in the control simulations never become fully optically revealed, although they do become substantially brighter over the course of the simulations, due to unrestrained accretion. Their apparent brightness can vary by several magnitudes on timescales of order $10^{5}$yr, due to non--steady flows of gas into the central regions of the clouds. In the dual--feedback simulations, we find that the timescales on which the clusters emerge from their veiling gas and become optically visible depends on the structure of the gas in the clouds and locate of the stars within it, but is short, generally 0.1--0.5Myr and never more than 1Myr. This timescale is substantially shorter than that on which the clouds are being disrupted. The times in which clouds I, J and UQ have half of their mass unbound are $>$2Myr. Although many of the stars in these simulations remain in regions of high extinction, the most massive and brightest objects which dominate the visible luminosity are very quick to disperse their surrounding material and become visible. This implies that the timescale for cloud dispersal and for the optical emergence of embedded clusters are not necessarily strongly connected.\\
\section{Conclusions}
We have evaluated the combined effects of photoionisation and wind feedback from O--type stars on the young embedded clusters formed in model turbulent GMCs of a range of masses and radii. We showed in previous work \citep[][]{2014MNRAS.442..694D} that the fraction of gas expelled from the clouds was strongly dependent on the cloud escape velocities, but here we concentrated on the effects of feedback on the clusters, the ongoing star formation process and the interplay between gas and stars. Our conclusions are outlined below:\\
{\indent (i) In the control simulations without feedback, star formation proceeds at the freefall rate in the dense gas, so that the depletion time of this material corresponds to its freefall time. In the feedback simulations the star formation rate becomes decoupled from the freefall time in the dense gas, sometimes being over an order of magnitude slower.\\
\indent (ii) The manner in which the star--formation--rate/dense-gas--mass relation is broken does not involve very large changes in the star formation rate, but instead is due to the production of dense but non--star--forming gas by feedback. Feedback expels dense gas from the clouds' potential wells, and generates dense gas outside the potential wells which is less able to form stars, but only affects the star formation rates by factors of at most two. If regions where star formation is occurring within clouds are spatially correlated with dense gas, the relationship still appears to be strong, as shown in Figure \ref{fig:sfr_sigma}, since star formation always takes place in dense gas. Instead, feedback creates large quantities of dense gas which fail to form stars.\\}
\indent (iii) The increase in dense gas but the similar or lower star formation rates in dual--feedback simulations relative to control simulations is due to the fact that most dense gas in the feedback calculations resides in the outskirts of the clouds where expanding bubbles have swept up low--density quiescent gas. The dense gas in these runs is not to be found in the depths of the clouds' potential wells which are, in the control simulations, the principal star--formation engines, serving to concentrate the dense gas further and preventing it from dispersing. Secondarily, the expulsion of gas from the potential wells and suppression of accretion flows feeding the clusters in the feedback simulations reduces the depths of the potential wells relative to those in the control calculations.\\
\indent (iv) Correlating \emph{star formation rates} with gas surface density does not reveal strong differences between the feedback and control simulations. These quantities show a strong relationship in both cases. Conversely, correlating \emph{stellar surface density} and gas surface density does reveal such differences. The separation of stars and dense gas caused by the clearing out of the central clusters and the generation of non--star--forming dense gas in the feedback simulations produces markedly different behaviour from the control simulations, where stars and dense gas remain correlated.\\
\indent (v) Simple modelling of the observed emergence of the clusters in the near--UV and optical bands implies that the apparent magnitudes of the clusters depend somewhat on the three--dimensional structure of the gas and the stellar system. The brightnesses of the clusters do not necessarily increase monotonically with time due to non--steady gas motions, particularly accretion flows in the control simulations. In the feedback simulations, the timescales on which the clusters become optically revealed can be substantially shorter than the timescales on which the host clouds are becoming unbound, and can also vary substantially with viewing angle due to anisotropy in the intracluster gas. Magnitude--limited surveys may therefore miss substantial numbers of clusters simply because of their orientation, and it may not be advisable to try to estimate the dispersion timescale by attempting to infer the revelation timescale observationally.\\
\indent (vi) Despite large quantities of damage being done to some clouds in terms of gas unbound or expelled, the mass or number fractions of \emph{stars} unbound by feedback are very modest and do not correlate very strongly with the amounts of gas lost. The cause of this discrepancy is likely to be the strong variations in local star formation efficiency, which approaches extremely high values in many of the clusters or sub clusters. The stars are often situated in stellar--dominated potentials in or close to virial equilibrium before the onset of feedback, so that the stellar systems are largely immune to the destruction of the surrounding clouds.\\

\section{Acknowledgements}
We thank the anonymous referee for stimulating comments and suggestions which improved the paper substantially. This research was supported by the DFG cluster of excellence `Origin and Structure of the Universe' (JED, BE). {We thank Nate Bastian for making available to us the stellar HST colours.}

\bibliography{myrefs}

\label{lastpage}

\end{document}